\documentclass[a4paper]{aa}
\usepackage{graphicx}
\usepackage{txfonts}  

\newcommand{\eu}[5]{\mbox{$#1\,^#2{\rm #3}^{#4}_{\rm #5}$}}

\newcommand{\Teff}{T_{\rm eff}}
\newcommand{\Eexc}{$E_{\rm exc}$}

\newcommand{\Vmic}{$V_{\rm mic}$}
\newcommand{\Vmac}{$V_{\rm mac}$}
\newcommand{\eps}[1]{\log\varepsilon_{\rm #1}}
\newcommand{\kH}{$S_{\rm H}$}
\newcommand{\logg}{\rm log~ g}
\newcommand{\bs}{$\langle B \rangle$}
\newcommand{\kms}{km\,s$^{-1}$}
\def\vs{$v_{\rm e}\sin i$}

\begin{document}

\title{Non-LTE line formation for \ion{Pr}{ii}\, and \ion{Pr}{iii}\, in A and Ap stars}

\author{L. Mashonkina\inst{1,2} \and T. Ryabchikova\inst{2,3} \and A. Ryabtsev\inst{4} 
\and R. Kildiyarova\inst{4}}

\offprints{L. Mashonkina}

\institute{Institut f\"ur Astronomie und Astrophysik der Universit\"at M\"unchen, Scheinerstr. 1, 81679 M\"unchen, Germany\\
\email{lyuda@usm.lmu.de} 
\and Institute of Astronomy, Russian Academy of Sciences, Pyatnitskaya 48, 119017 Moscow, Russia\\
\email{lima@inasan.ru} 
\and Institute for Astronomy, University of Vienna, T\"urkenschanzstrasse 17, A-1180 Vienna, Austria 
\and Institute of Spectroscopy, Russian Academy of Sciences, 142190, Troitsk, Moscow region, Russia 
}
\date{Received  / Accepted }

\abstract{}
{Non-local thermodynamical equilibrium (non-LTE) line formation for
singly-ionized and doubly-ionized praseodymium is considered
through a range of effective temperatures between 7250~K and 9500~K. We evaluate the influence of
departures from LTE on Pr abundance determinations and determine a distribution of the Pr abundance in the atmosphere of the roAp star HD\,24712 from non-LTE analysis of the \ion{Pr}{ii} and \ion{Pr}{iii} lines.}
{A comprehensive model atom for \ion{Pr}{ii/iii}\, is presented based on the measured and the predicted energy levels, in total, 6708 levels of \ion{Pr}{ii}\, and  \ion{Pr}{iii} and the ground state of \ion{Pr}{iv}. Calculations of the \ion{Pr}{ii}\, energy levels and oscillator strengths for the transitions in \ion{Pr}{ii}\, and \ion{Pr}{iii}\, are described.}
{The dependence of non-LTE effects on the atmospheric parameters is discussed. At $\Teff \ge$ 8000~K departures from LTE  lead to overionization of \ion{Pr}{ii}\, and, therefore, to systematically depleted total absorption in the line and positive abundance corrections.
 At the lower temperatures, different lines of \ion{Pr}{ii}\, may be either weakened or amplified depending on the line strength. The non-LTE effects strengthen the \ion{Pr}{iii}\, lines and lead to negative abundance corrections. Non-LTE corrections grow with effective temperature for the \ion{Pr}{ii}\, lines, and, in contrast, they decline for the \ion{Pr}{iii}\, lines. The \ion{Pr}{ii/iii} model atom is applied to determine the Pr abundance in the atmosphere of the roAp star HD\,24712 from the lines of two ionization stages. In the chemically uniform atmosphere with [Pr/H] = 3, the departures from LTE  may explain only small part (approximately 0.3~dex) of the difference between the LTE abundances derived from the \ion{Pr}{ii} and \ion{Pr}{iii} lines ($\simeq$2~dex). We find that the lines of both ionization stages are described for the vertical 
distribution of the praseodymium where the Pr enriched layer with [Pr/H] $\ge$ 4 exists in the outer atmosphere at $\log \tau_{5000} < -4$. 
 The departures from LTE for \ion{Pr}{ii/iii} are strong in the stratified atmosphere and have the opposite sign for the \ion{Pr}{ii} and \ion{Pr}{iii} lines. The praseodymium stratification analysis of roAp stars has to be performed based on non-LTE line formation.
Using the revised partition function of \ion{Pr}{ii} and experimental transition probabilities, we determine the solar non-LTE abundance of Pr as $\log\rm {(Pr/H)}_\odot=-11.15\pm0.08$.}
{}


\maketitle

\section{Introduction}

Classical LTE analysis finds a great violation of the ionization
equilibrium between the second and the first ions of the
rare-earth elements (REE) in rapidly oscillating chemically
peculiar (roAp) stars (Cowley \& Bord \cite{cowley98},
Cowley~et~al. \cite{cowley2000}, Gelbmann~et~al. \cite{gelbman},
Ryabchikova~et~al. \cite{RSHWH00}, Kochukhov \cite{K03},
Ryabchikova~et~al. \cite{ryab2001}). For the sample of 26
stars, Ryabchikova et~al. (\cite{ryab2001}) show that a
discrepancy between the abundances derived from the lines of singly ionized and doubly 
ionized atoms of neodymium and praseodymium typically exceeds 1.5~dex in roAp stars, while 
it is substantially smaller if exists in non-pulsating Ap stars
(see also Kato \cite{Kato03}; Ryabchikova et~al. \cite{RRKB06} for hotter Ap stars HD~170973 and HD~144897).   
This rules out errors in the oscillator strengths as a possible reason for the abundance difference observed in roAp stars.

In our previous paper (Mashonkina et~al. \cite{mash_nd_ap}), we
investigate the \ion{Nd}{ii}\, and \ion{Nd}{iii}\, spectra in two
roAp stars, $\gamma$~Equ and HD\,24712, based on non-local
thermodynamic equilibrium (non-LTE) line formation and show that
the non-LTE effects may explain only 0.5~dex of the difference
between the LTE abundances derived from the \ion{Nd}{ii}\, and
\ion{Nd}{iii}\, lines and not the 1.5 -- 2.0~dex observed in the
these stars. Mashonkina et~al. come back, therefore, to the
assumption of Ryabchikova et~al. (\cite{ryabchik02}) that the Nd
anomaly observed in $\gamma$~Equ is caused by a stratified Nd
distribution with the accumulation of the element in the uppermost
atmospheric layers, above $\log\tau_{5000}=-8$ according to the
LTE analysis. It was found that the non-LTE effects for the
\ion{Nd}{ii}\, and \ion{Nd}{iii}\, lines are very strong in the
stratified atmosphere, and they result in significant shifting the Nd enriched layer downward compared to the location determined in the LTE analysis.
The required Nd overabundance in the layer is [Nd/H] = 4 at $\log \tau_{5000} <
-3.5$ for $\gamma$~Equ and [Nd/H] = 4.5 at $\log \tau_{5000} <
-4.5$ for HD\,24712. 

The present paper continues to investigate the rare-earth elements in stellar atmospheres based on the non-LTE line formation and is devoted to the praseodymium. 
We study the
statistical equilibrium (SE) of singly ionized and doubly ionized praseodymium, \ion{Pr}{ii}\, and \ion{Pr}{iii}, through a range of effective temperatures 
between 7250~K and 9500~K, evaluate the influence of departures from LTE on Pr abundance determinations in the Sun, A and Ap type stars, consider the 
non-LTE effects for \ion{Pr}{ii/iii}\, in the atmosphere with non-uniform vertical distribution of Pr, and finally determine empirically a stratification of 
Pr in the atmosphere of the roAp star HD\,24712 from the non-LTE analysis of the \ion{Pr}{ii}\, and \ion{Pr}{iii}\, lines.

The paper is organized as follows. In Sect.\,\ref{atom},  an extensive model atom for \ion{Pr}{ii/iii} is introduced and theoretical calculations of the \ion{Pr}{ii}\, 
atomic structure and transition probabilities for \ion{Pr}{ii}\, and \ion{Pr}{iii} are presented.
The programs and atmospheric models used in the line formation calculations are described in Sect.\,\ref{Programs}. Section~\ref{departures} investigates the 
departures from LTE for \ion{Pr}{ii/iii}\, in the model atmospheres with homogeneous and stratified distribution of Pr. The non-LTE abundance corrections for 
the selected lines of \ion{Pr}{iii}\, 
and \ion{Pr}{ii}\, are given there depending on effective temperature. 
The solar Pr abundance is revised in
Sect.\,\ref{sun} based on the improved partition function of \ion{Pr}{ii}. In Sect.~\ref{obs}, we determine the Pr
abundance distribution in the atmosphere of the roAp star
HD\,24712 from the non-LTE analysis of the \ion{Pr}{ii}\, and
\ion{Pr}{iii}\, lines and discuss the influence of the
uncertainties of atomic parameters on non-LTE modelling and final
results. Our conclusions and recommendations are given 
in Sect.~\ref{end}.

\section{Model atom of \ion{Pr}{ii}-\ion{Pr}{iii}} \label{atom}

Model atom provides the necessary atomic input data to specify the
SE equations and the opacities/emissivities for radiative transfer calculations.

\subsection{Energy levels}\label{levels}

The lower levels in singly ionized Pr belong to the $4f^3 6s$
configuration with the ($^4$I)$^5$I ground term. Laboratory measurements (Ginibre \cite{ginibre1, ginibre2}, Furman et~al. \cite{furman}, Ivarsson et~al. \cite{ILW}) give 330 energy levels of \ion{Pr}{ii}\, with an excitation energy \Eexc\, $\le$ 5.3~eV. Most known levels belong to singlet, triplet, and quintet terms of the $4f^3~nl$ ($nl = 6s, 5d, 6p$) and $4f^2 5d~nl$ ($nl = 5d, 6s, 6p$) electronic configurations. For some of the known energy levels, only the total angular momentum and assignment to an electronic configuration were given. The highest known levels of \ion{Pr}{ii}\, are separated by more than 5~eV from the ground state of \ion{Pr}{iii}. The calculations with the Cowan (\cite{cowan}) code show that even below 5.3~eV there are many unidentified energy levels. With such an incomplete term system we cannot get a realistic statistical equilibrium of the atom.

For the present study, we calculate the energy levels of \ion{Pr}{ii}\, using the Cowan code (Cowan \cite{cowan}). The measured levels of the odd $4f^3 6s + 4f^3 5d + 4f^2 5d6p$ configurations have been fitted taking into account the interactions with the $4f^2 6s6p$, $4f 5d^3$, $4f 5d^2 6s$, and $4f 5d 6s^2$ configurations. The $4f^2 5d^2$, $4f^2 5d6s$, $4f^2 6s^2$, $4f^3 6p$, $4f^4$, $4f^2 6p^2$, and $4f 5d^2 6p$ configurations are included in the fitted matrix of the even system. The fitting results in a standard deviation of the calculated from experimental levels of 86~cm$^{-1}$ for 129 odd and 72~cm$^{-1}$ for 201 even levels. To get highly excited levels the energy structures of the even $4f^3(7p-9p)$ and $4f^3(5f-6f)$ configurations and the odd $4f^3(7s-9s)$, $4f^3(6d-8d)$, $4f^35g$, and $4f^25d5f$ configurations are calculated. In all unknown configurations, the average energies where Brewer (\cite{Brewer}) predictions are absent and the energy parameters of the $4f^2$ and $4f^3$ cores are scaled similarly to the known configurations, the other parameters being scaled by a factor 0.75 with respect to the corresponding Hartree - Fock values. As a result, a nearly complete set of the levels below 70000~cm$^{-1}$ (8.7~eV) has been obtained consisting of 3938 energies.
All the known levels and the predicted levels with \Eexc\, $\le$ 10.04~eV, in total 6539 levels of \ion{Pr}{ii}, are used to construct the model atom. They are shown in Figs.\,\ref{pr2_even} and 
\ref{pr2_odd} (Online only) for the even and odd levels, correspondingly. 
The calculated high excitation levels provide the close collisional coupling of \ion{Pr}{ii}\, to the continuum electron reservoir. 

It is worth noting that the calculations lead to a significantly larger partition function of \ion{Pr}{ii}\, compared to that based on the laboratory levels only. 
This is illustrated in Table\,\ref{pf}, where the partition function of \ion{Pr}{ii} is calculated by its definition for a temperature range 3000~K -- 20000~K 
using the measured energy levels from Ginibre (\cite{ginibre1, ginibre2}) and using the levels predicted in this study.
In Sect.\,\ref{sun}, we calculate the effect of the revised partition function on the Pr abundance determined from the solar \ion{Pr}{ii}\, lines.

\begin{table}
\begin{center}
\caption{Partition functions for \ion{Pr}{ii} and \ion{Pr}{iii}.}\label{pf}
\begin{tabular}{c|cc|c}
\hline\noalign{\smallskip}
 Temperature& \multicolumn{2}{c|} {\ion{Pr}{ii}} & \ion{Pr}{iii}    \\
\cline{2-3}
   (K)     & Ginibre (1989) & this study & this study \\  
\hline
 3000 &~~55 &~~58&~22  \\
 4000 &~~91 &~101&~28  \\
 5000 &~138 &~166&~34  \\
 6000 &~192 &~251&~43  \\
 7000 &~252 &~356&~54  \\
 8000 &~315 &~478&~66  \\
 9000 &~380 &~617&~81  \\
10000 &~446 &~770&~97  \\
11000 &~511 &~937&114  \\
12000 &~576 &1117&132  \\
13000 &~639 &1309&151  \\
14000 &~701 &1514&170  \\
15000 &~762 &1732&190  \\
16000 &~820 &1962&211  \\
17000 &~878 &2205&231  \\
18000 &~933 &2460&252  \\
19000 &~986 &2727&273  \\
20000 &1038 &3007&293  \\
\hline
\end{tabular}
\end{center}
\end{table}

For \ion{Pr}{iii}, the laboratory measurements resulted in 593 energy levels (Martin et~al. \cite{NIST}, Palmeri et~al. \cite{Palmeri}) with an excitation energy up to 17.5~eV. 
In the range of stellar parameters we are concerned with, there is no need to include the highly excited levels of \ion{Pr}{iii}\, in the model atom. They play a minor role in population and depopulation of \ion{Pr}{iii}, because the next ionization stage Pr\,{\sc iv}\ represents a negligible fraction of Pr abundance. We use, therefore, the levels of the odd $4f^3$ and the even $4f^25d$ and $4f^26s$ electronic configurations with \Eexc\, $\le$ 6.7~eV. The term structure is shown in Fig.\,\ref{pr3} (Online only). 
The contribution of the \ion{Pr}{iii}\, energy levels omitted in the final model atom to the \ion{Pr}{iii}\, partition function is less than 0.01\%\ at $T_{\rm e}$ = 7250~K. 


Levels of the same parity with small energy differences were combined into a single level.  The final model atom includes 294 combined levels of
\ion{Pr}{ii}, 54 combined levels of \ion{Pr}{iii}, and the ground
state of \ion{Pr}{iv}. 

\onlfig{1}{
\begin{figure*}
\includegraphics[width=16cm]{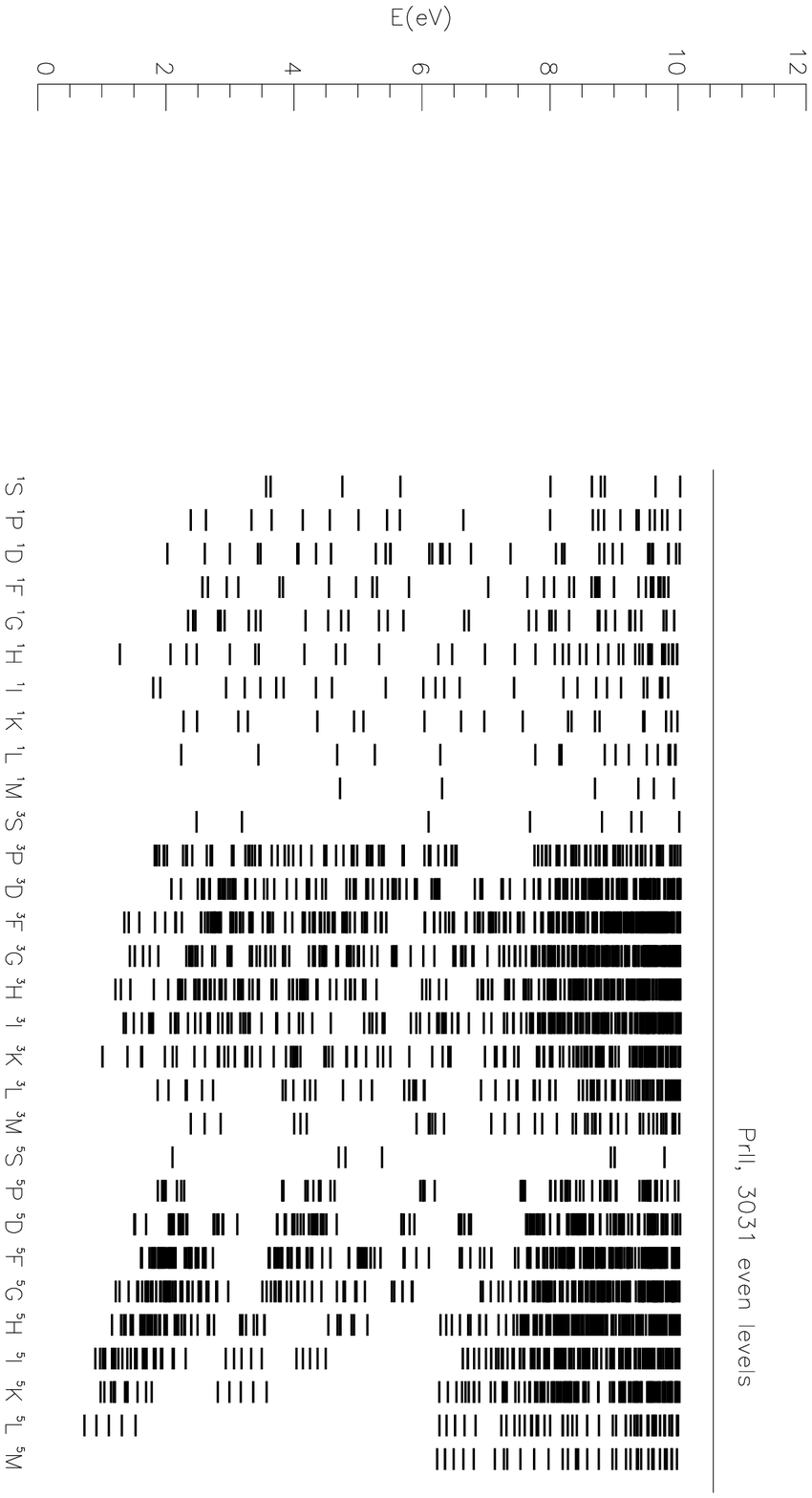}
\caption {The \ion{Pr}{ii} even term structure.} \label{pr2_even}
\end{figure*}
}

\onlfig{2}{
\begin{figure*}
\includegraphics[width=16cm]{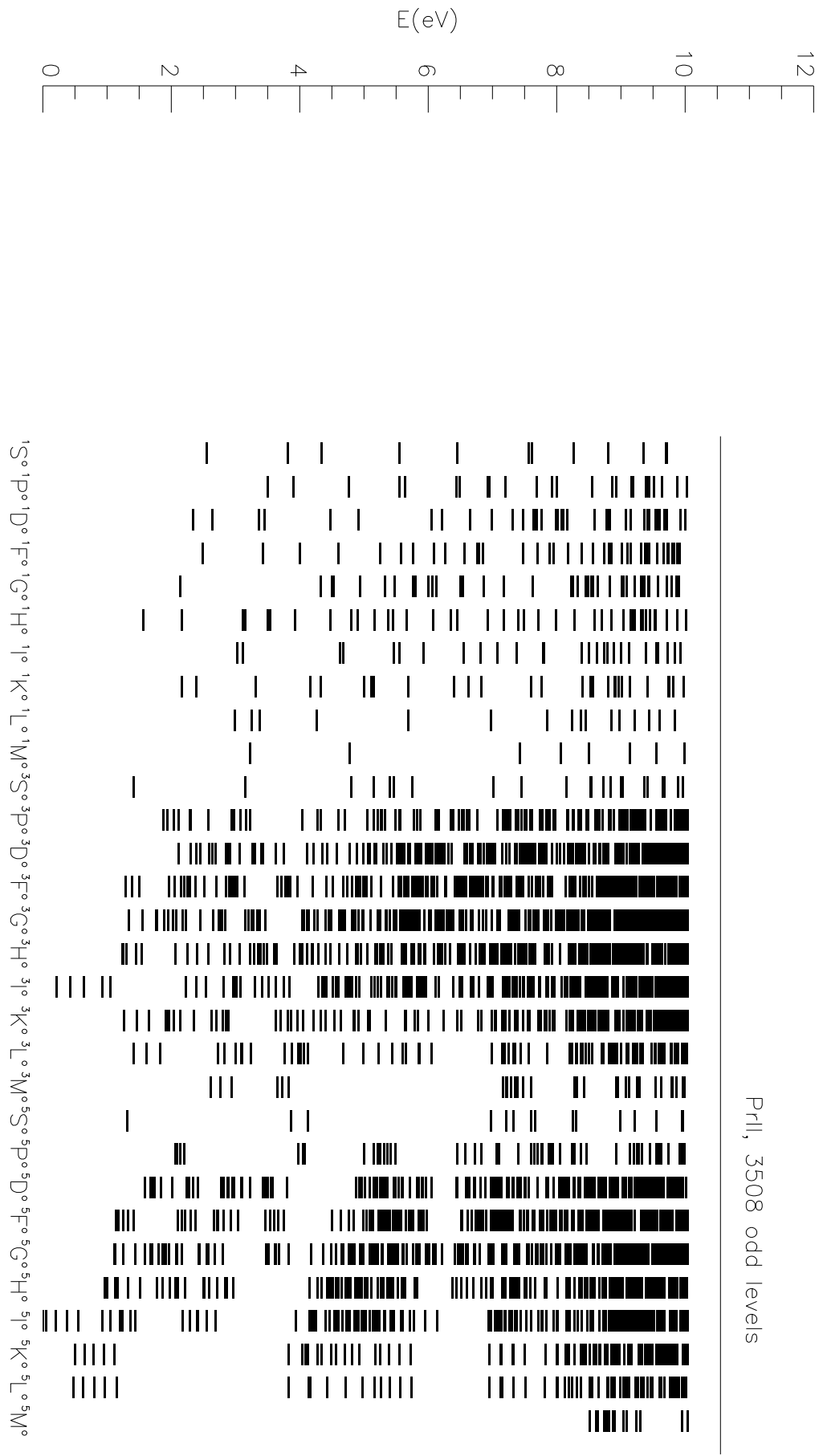}
\caption {The \ion{Pr}{ii} odd term structure.} \label{pr2_odd}
\end{figure*}
}

\onlfig{3}{
\begin{figure*}
\includegraphics[width=16cm]{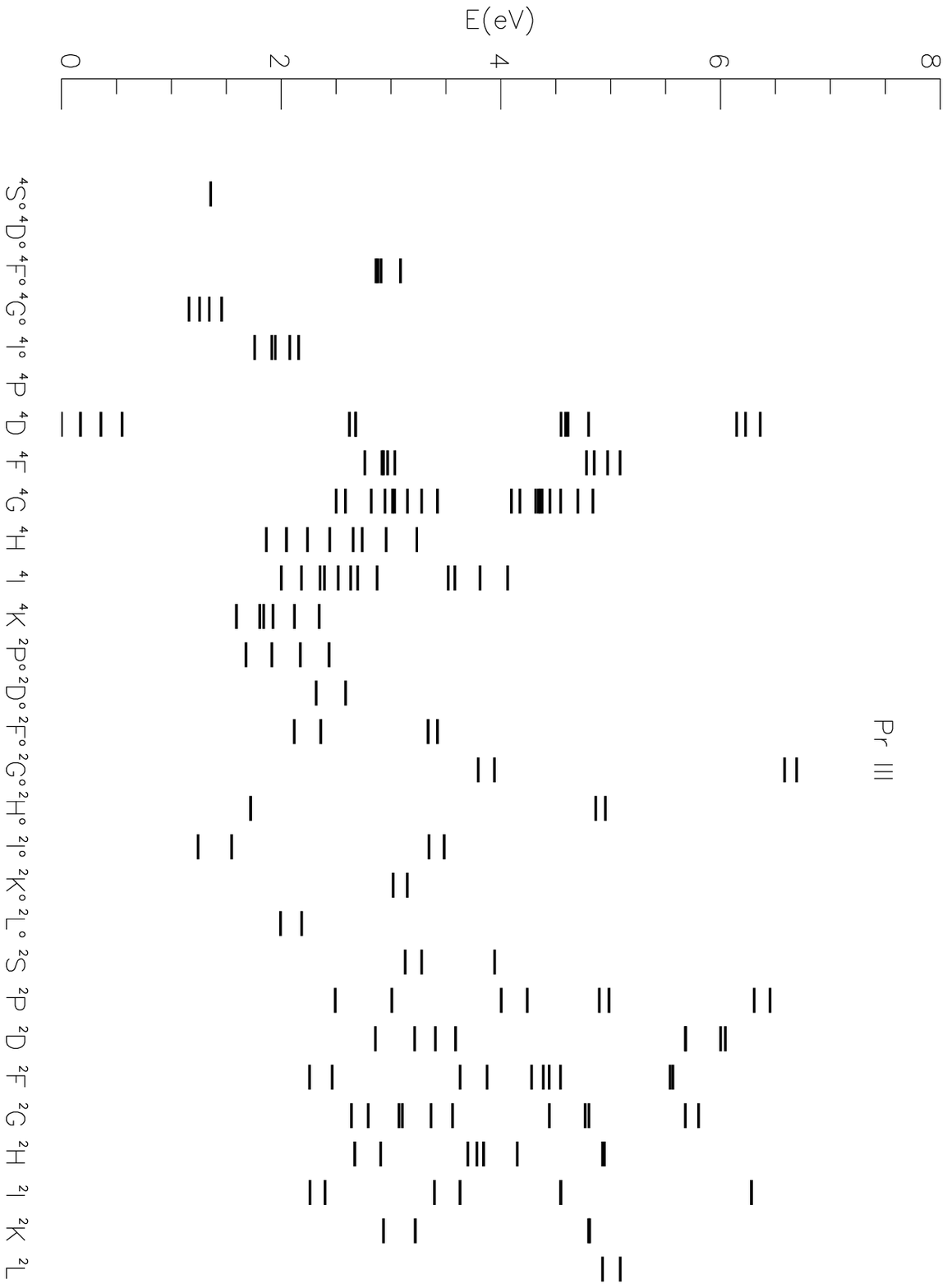}
\caption {The \ion{Pr}{iii} model atom.} \label{pr3}
\end{figure*}
}

\subsection{Radiative data}
\label{trans}

The 15788 and 392 radiative bound-bound transitions in \ion{Pr}{ii}\, and \ion{Pr}{iii}, respectively, are included in SE calculations. Oscillator strengths $f_{ij}$ based on laboratory measurements of Lage \& Whaling (\cite{LW}) or estimated from the observed line
intensities of Meggers et~al. (\cite{MCS}) are available only for 448 transitions in \ion{Pr}{ii}. They come from the Kurucz \& Bell (\cite{Kur94b}) linelist and are accessible via the Vienna Atomic Line Data Base ({\sc vald}, Kupka et~al. \cite{vald}).

For the majority of transitions in \ion{Pr}{ii}\, and all transitions in \ion{Pr}{iii}, we rely, therefore, on oscillator strengths computed in the present study. 
Calculations for \ion{Pr}{ii} are based on the wave functions obtained in the fittings of the energy levels.
All the Hartree-Fock transition integrals are scaled by a factor 0.85. The calculated lifetimes have been compared with the recent accurate measurements of Scholl et~al. (\cite{Scholl}) and Bi\'emont et~al. (\cite{B2003}). 
A quite good agreement is found for the levels with leading contribution from the levels of the $4f^36p$ configuration: the ratio of the measured to calculated lifetime equals 1.1 with a standard deviation 0.3. For the levels with highly mixed wave functions, the corresponding quantity appears to be 0.7 with the standard deviation 0.5. Details of these calculations can be found on \verb|http://das101.isan.troitsk.ru/files/SPECTRA/Pr_II| and will be presented in a forthcoming paper.

 For \ion{Pr}{iii}, our approach follows that of Palmeri et~al. (\cite{Palmeri}). Fitting of the energy levels is performed taking into account the interactions in the odd $4f^3$ + $4f^26p$ + $4f5d^2$ + $4f5d6s$ + $4f6s^2$ +  $4f^25f$ + $5p^54f^4$ and the even $4f^25d$ + $4f^26d$ + $4f^26s$ + $5p^54f^35d$ + $5p^54f^36s$ + $4f5d6p$ + $4f6s6p$ + $4f5d5f$ complexes. {\em Ab initio} transition integrals are taken. For 7 levels with a more than 25\%\ contribution of the $4f^26p$ configuration, the average ratio of the measured (Bi\'emont et~al. \cite{B2001}) to calculated lifetimes is obtained to be $1.17 \pm 0.25$.  An exception is the $J = 9/2$ level at 62535.6~cm$^{-1}$ with main contribution from the $4f5d^2$ configuration and about 11\%\ admixture of the $4f^26p$ configuration. For this level, the lifetime ratio appears to be 3.1. 
Details of the \ion{Pr}{iii}\, calculations relevant to this article can be found on the website \verb|http://das101.isan.troitsk.ru/files/SPECTRA/Pr_III|.

The photoionization cross-sections $\sigma_{\rm ph}$ are computed using the hydrogen approximation because no accurate data is available for the \ion{Pr}{ii}\, and \ion{Pr}{iii}\,
levels. We assume that the
photoionization from any \ion{Pr}{ii}\, level ends in the ground
state of \ion{Pr}{iii}, \eu{4f^3}{4}{I}{\circ}{9/2}. To take into
account the photoionization to the remaining levels of the
\eu{4f^3}{4}{I}{\circ}{} term we multiply $\sigma_{\rm ph}$ by the
ratio
$g$(\eu{4f^3}{4}{I}{\circ}{})/$g$(\eu{4f^3}{4}{I}{\circ}{9/2})
$\simeq$ 5. The photoionization from the $4f^25dnl$ levels ends
in the excited \ion{Pr}{iii}\, $4f^25d$ levels, however, we neglect this for the following reason. 
Replacing the real ending state on the \ion{Pr}{iii}\, ground
state makes each $4f^25dnl$ level to be easier ionized. Its
ionization energy is reduced by more than 2~eV, and the photoionization rate is
overestimated, in particular, for the levels with a
threshold in the ultraviolet (UV). At the same time, the photoionization rates of the $4f^25dnl$ levels are underestimated due to  
ignoring multiple channels for their ionization. We cannot
estimate net effect due to absence of accurate atomic data. 

\subsection{Collisional data}

The calculations of electron impact
excitation and ionization rates rely on theoretical approximations. We use the formula of van Regemorter (\cite{Reg}) for the allowed transitions and assume that the effective collision strength $\Upsilon$ = 1 for the forbidden transitions. Electron impact ionization cross-sections are
computed according to Drawin (\cite{Draw}).

The effects of the uncertainties of the used photoionization cross-sections
and collisional rates on the final results are described in
Sect.~\ref{uncertainty}.

\section{Programs and model atmospheres}\label{Programs}

The radiative transfer and the statistical equilibrium equations are solved 
with a revised version
of the DETAIL program (Butler \& Giddings \cite{detail}) using the
accelerated lambda iteration following the extremely efficient
method described by Rybicki \& Hummer (\cite{rh91}, \cite{rh92}). 
Background opacities include the important bound-free and free-free transitions of hydrogen, helium, and the most abundant metals, the Rayleigh scattering, the Thomson scattering, the hydrogen lines, the quasi-molecular Lyman $\alpha$ satellites due to H-H and H-H$^+$ collisions (Allard et~al.\cite{allard98}), and line opacity calculated with the line lists made available by Kurucz \& Bell (\cite{Kur94b}). The background opacities are sampled on a random grid of 4500 frequencies, to which are added the frequencies of the line profiles. The final non-LTE line formation program samples roughly 150 000 wavelengths between 500\AA\ and 80 000\AA.  

The obtained non-LTE and LTE level populations are used to compute the emergent flux, line profiles, and equivalent widths with the code LINEC. The investigated lines and transitions of \ion{Pr}{ii}\, and \ion{Pr}{iii}\, are
listed in Table\,\ref{lines}. The wavelengths of \ion{Pr}{ii}\, lines are taken from Ginibre (\cite{Ginibre1990}). Ritz wavelengths are used for the \ion{Pr}{iii}\, lines.

 Hyperfine-structure (HFS) affecting the \ion{Pr}{ii} lines is explicitly calculated with the HFS constants given by Ginibre (\cite{ginibre1}), 
if available. No HFS data exist in the literature for the \ion{Pr}{iii} lines. Ivarsson et~al. (\cite{ILW}, ILW)  could not resolve HFS in any of the
accurately measured \ion{Pr}{iii} lines, which means that HFS is not large, although some lines show asymmetric profiles with half-widths exceeded
the half-width of the resolved HFS components in \ion{Pr}{ii}. The three from eight common \ion{Pr}{iii} lines are practically HFS 
unaffected according to the ILW laboratory analysis. Half-widths of other three lines are three to four times larger compared to those for the HFS 
unaffected lines, and they are comparable with the thermal half-widths of the \ion{Pr}{iii} lines in the atmosphere of HD~24712. The thermal half-width of the Pr lines corresponds to approximately 1~\kms. Our simulation of HFS affecting the \ion{Pr}{iii} 5998 and 6053 lines with the strongest HFS broadening according to the measurements of Ivarsson et~al. (\cite{ILW}) leads to an increase of the theoretical equivalent widths, by 16~m\AA\ and 17~m\AA, respectively, that transforms to --0.20~dex and --0.24~dex decrease of the element abundance derived from these lines. Further analysis is performed not accounting for HFS for the \ion{Pr}{iii} lines. We show in Sect.~\ref{pr_hd24712}, that two abovementioned lines give the results consistent with those for the \ion{Pr}{iii} lines with weak HFS broadening.

We use homogeneous blanketed model atmospheres. 
The small grid of models with $\Teff$ ranging between 7500~K and 9500~K with a step of 500~K, $\logg$ = 4, and the solar chemical composition has
been calculated with the MAFAGS code (Fuhrmann et~al.
\cite{Fuhr1}) that treats line-blanketing using the opacity
distribution functions (ODF). The original ODF tables of Kurucz (\cite{Kur94}) were scaled by $-0.16$~dex to put the iron opacity
calculated by Kurucz with $\eps{Fe} = 7.67$ into correspondence
with a value $\eps{Fe} = 7.51$ which we believe to be the best representation of the solar mixture. We refer to abundances on the usual scale where $\eps{H} = 12$.

For HD\,24712 ($\Teff$ = 7250~K, $\logg$ = 4.3, [M/H] = 0 according to Ryabchikova et~al. \cite{RYABCHIK97}), the model atmosphere
has been computed by Frank Grupp with the MAFAGS-OS code (Grupp
\cite{grupp}). It is based on up-to-date continuous opacities and
includes the effects of line-blanketing through opacity sampling.

 HD\,24712 possesses a magnetic field with a mean magnetic field modulus \bs\ changing from 2.5~kG at the magnetic minimum to $\sim$3.1--3.3~kG at the
magnetic maximum (Ryabchikova et~al. \cite{Ryabchikova2007a}). We ignore the influence of the magnetic field on atmospheric structure, based on the 
results of Kochukhov et~al. (\cite{Ketal05}) who have shown that the difference in temperature and gas pressure distributions between magnetic and 
non-magnetic model atmospheres with $\Teff$ = 8000~K does not exceed 30~K and 6\%, respectively, for field strengths up to 5~kG. 
However, splitting of the spectral lines in the magnetic field is taken into account in LTE abundance analysis. Magnetic spectrum synthesis is performed 
with the help of {\sc SYNTHMAG} code (Kochukhov \cite{synth3}).

There are observational evidences for non-uniform element distribution in the atmosphere of HD\,24712. The neodymium is strongly enhanced in the uppermost 
atmospheric layers according to Mashonkina et~al. (\cite{mash_nd_ap}). Ryabchikova et al. (\cite{Ca-IS}) and Ryabchikova (\cite{RYABCHIK2008}) show that Ca, Si, Cr, Fe, Sr, and Ba are concentrated in deep atmospheric layers. One believes that radiatively driven diffusion is one of the main processes responsible for these inhomogeneities. The self-consistent diffusion 
models predict the vertical distributions of Mg, Si, Ca, Ti, Fe which qualitatively reproduce the corresponding element stratifications found empirically 
for some Ap stars (Alecian \& Stift \cite{alecian2007}, LeBlanc \& Monin \cite{leblanc}). But no theoretical predictions are available for the REE due 
to an incompleteness of atomic data on energy levels and transition probabilities. Stratified distribution of chemical elements in the atmosphere can 
influence the atmospheric structure. Modelling chemically non-uniform stellar atmospheres based on the abundance gradients determined empirically is in progress (Shulyak, 2008, private communication). 
 Preliminary results of the iterative procedure for HD~24714 predict the slight change
in the model parameters ($\Teff$ = 7250~K and $\logg$ = 4.1) and  
lead to only minor changes in our results on Pr analysis presented in Sect.\,\ref{obs}.

\section{Departures from LTE for \ion{Pr}{ii}-\ion{Pr}{iii}}\label{departures}

In line formation layers of the atmospheres with $\Teff$ between
7250~K and 8000~K, number density of \ion{Pr}{ii} is larger compared to that for \ion{Pr}{iii} (see Fig.\,\ref{ion7500}). However, \ion{Pr}{ii} drops rapidly in
the hotter atmospheres. Non-LTE calculations show that the main
non-LTE effect for \ion{Pr}{ii}\, is overionization caused by a
super-thermal radiation of non-local origin near the thresholds of the 
$4f^36p$ levels with \Eexc\,= 3~eV -- 4~eV ($\lambda_{\rm
thr} =$ 1600\AA\, to 1850\AA). Photoionization is able to drain
the populations of these 
levels in the model with the lowest temperature, $\Teff$ = 7250~K, and
the effect is strengthened with $\Teff$ increasing. The population
loss is redistributed over many levels, producing overall
depletion of the first ionization stage in line formation layers.
In contrast, photoionization of the \ion{Pr}{iii}\, levels is
inefficient 
due to very low stellar fluxes in the far ultraviolet ($\lambda <$ 900\AA),
where the ionization edges of the \ion{Pr}{iii}\, ground state and the 
low excitation levels are located. At $\Teff \leq$ 8000~K, overionization of \ion{Pr}{ii}\,
leads to overpopulation of \ion{Pr}{iii}.  At the higher temperatures, 
\ion{Pr}{iii}\, represents the majority of the element and
preserves the thermodynamic equilibrium (TE) total number density.

We find that the mechanisms driving departures from LTE for
\ion{Pr}{ii/iii}\, are similar in the atmospheres with uniform and
stratified distribution of praseodymium, however, the magnitude of the
effect is different due to a different location of the line
formation layers.

\subsection{The atmospheres with uniform distribution of praseodymium}

\begin{figure}
\begin{center}
\resizebox{88mm}{!}{\includegraphics{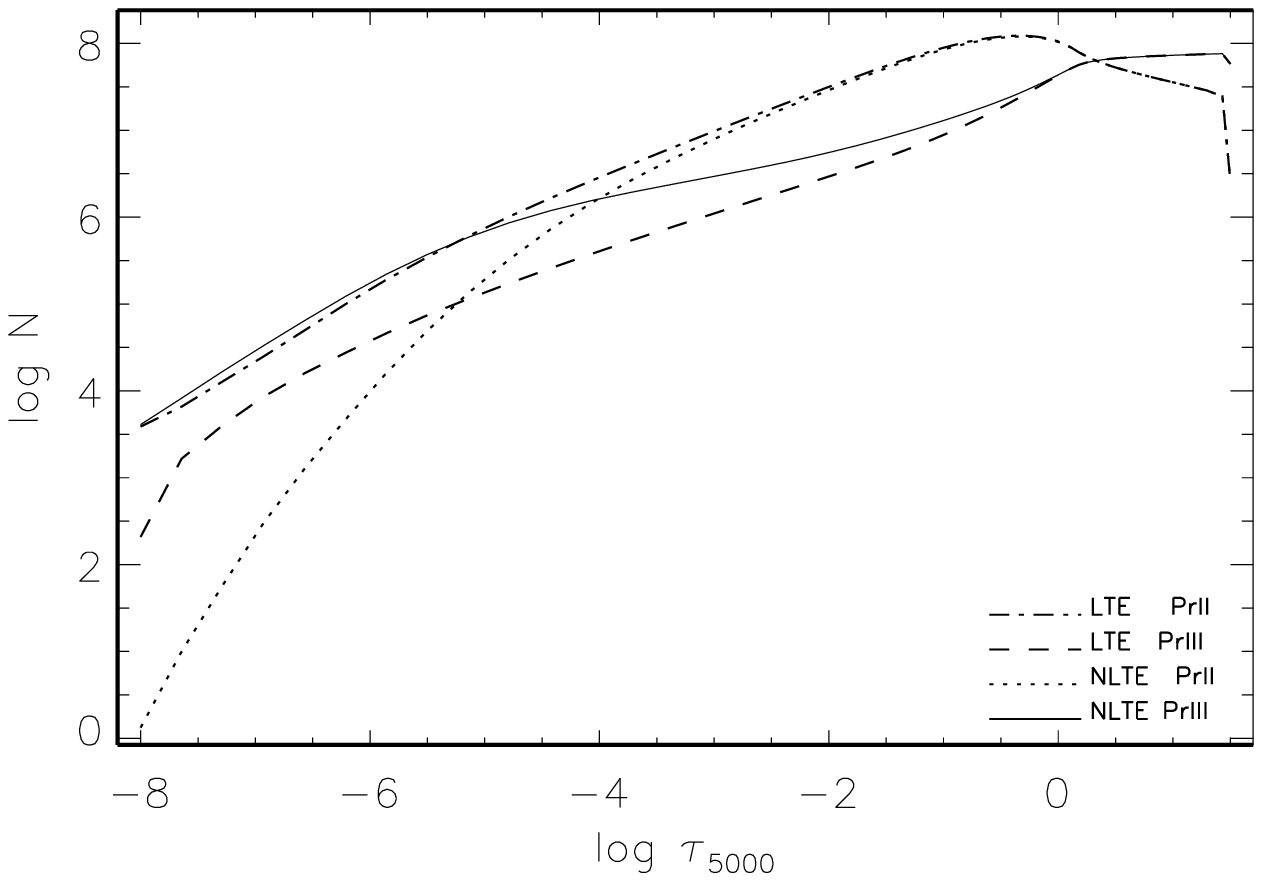}}

\vspace{-5mm} 
\resizebox{88mm}{!}{\includegraphics{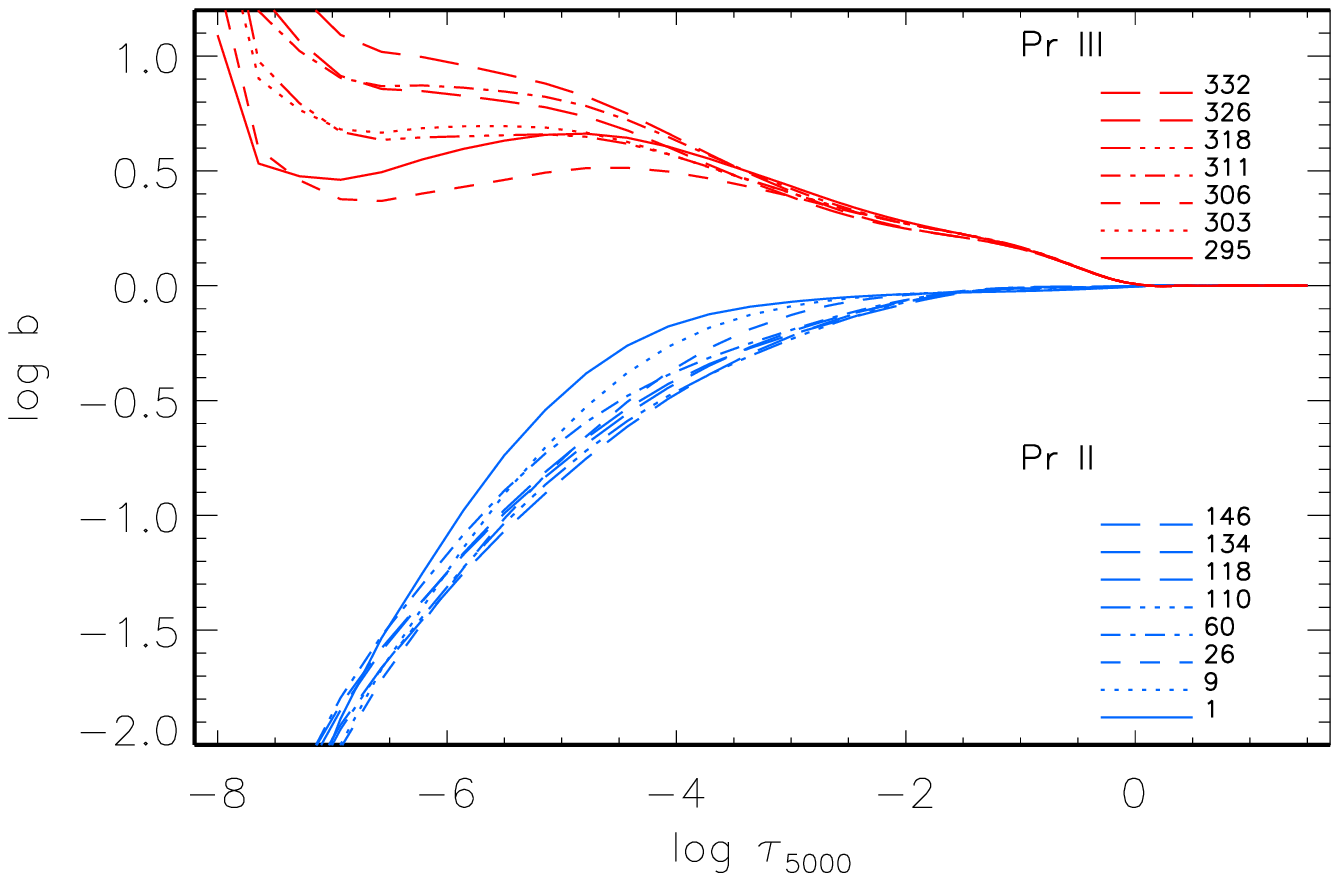}}
\caption[]{LTE and non-LTE total number densities of
\ion{Pr}{ii}\, and \ion{Pr}{iii}\, (top panel) and departure
coefficients for selected levels of \ion{Pr}{ii}\, and
\ion{Pr}{iii}\, (bottom panel) in the model atmosphere 7250/4.3/0.
Successive numbers of the levels in our model atom are quoted.
Everywhere in the atmosphere [Pr/H] = 3.} \label{ion7500}
\end{center}
\end{figure}

 In the observed spectrum of HD\,24712, 14 lines of \ion{Pr}{iii} are detected with a measured equivalent width, $W_\lambda$, of 17~m\AA\ and larger. 
In order to predict the equivalent widths of the \ion{Pr}{iii} lines at a detectable level
in the model, which represents the atmosphere of HD\,24712, we perform calculations with the praseodymium abundance [Pr/H] = 3. This value also characterizes the 
mean Pr abundance derived from all investigated \ion{Pr}{ii} and \ion{Pr}{iii} lines in HD\,24712.
Figure\,\ref{ion7500} shows the departure coefficients, $b_i = n_i^{\rm NLTE}/n_i^{\rm
LTE}$ of the selected levels of \ion{Pr}{ii}\, and \ion{Pr}{iii}\, as a function of continuum optical depth 
$\tau_{5000}$ at $\lambda = 5000$\AA\, 
in the model atmosphere with $\Teff$/$\logg$/[M/H] = 7250/4.3/0.
Here, $n_i^{\rm NLTE}$ and
$n_i^{\rm LTE}$ are the statistical equilibrium and TE
(Saha-Boltzmann) number densities, respectively. The medium becomes optically
thin for the ionizing radiation below 1850\AA\, far inside the
atmosphere, at $\log\tau_{5000}$ around 0. As a result, \ion{Pr}{iii}\, is overpopulated in line formation layers, at $\log\tau_{5000} < 0$. The \ion{Pr}{ii}\, levels with \Eexc $<$ 6~eV are strongly coupled to the \ion{Pr}{ii}\, ground state and to each other inside $\log\tau_{5000} \simeq -1.7$ where their departure coefficients are only slightly below 1. The outside layers become transparent for the radiation of many
\ion{Pr}{ii}\, lines arising between low-excitation terms (\Eexc $<$
2 eV) and intermediate-excitation terms (\Eexc\, = 2.7 - 4~eV).
The photon loss in these lines amplifies the underpopulation of
the upper levels (level numbers from 110 to 146 in
Fig.\,\ref{ion7500}) caused by enhanced photoionization.

\begin{table}
\caption{Investigated transitions in
\ion{Pr}{ii}\, and \ion{Pr}{iii}. } \label{lines}
\begin{tabular}{lclc}
\hline\noalign{\smallskip}
 $\lambda$ [\AA] & \Eexc (eV) & \multicolumn{2}{c}{Transition} \\
\hline
\multicolumn{4}{l}{ \ \ \ion{Pr}{ii}} \\
4222.93$^{1*}$ & 0.05 &\eu{4f^35d}{5}{I}{\circ}{5} - \eu{4f^36p}{5}{K}{}{6}  & ~~2 - 118$^2$ \\
4449.83$^{1*}$ & 0.20 &\eu{4f^35d}{5}{I}{\circ}{6} - \eu{4f^36p}{5}{K}{}{6}  & ~~3 - 118 \\
5002.44     & 0.80 &\eu{4f^35d}{5}{K}{\circ}{7} - \eu{4f^36p}{3}{K}{}{6}  & ~12 - 129 \\
5110.76$^*$ & 1.15 &\eu{4f^35d}{5}{L}{\circ}{10} - \eu{4f^36p}{5}{K}{}{9} & ~26 - 143 \\
5129.54$^*$ & 0.65 &\eu{4f^35d}{5}{K}{\circ}{6} - \eu{4f^36p}{5}{I}{}{5}  & ~10 - 121 \\
5135.14$^*$ & 0.95 &\eu{4f^35d}{5}{K}{\circ}{8} - \eu{4f^36p}{5}{K}{}{8}  & ~16 - 134 \\
5259.73$^{1*}$ & 0.63 &\eu{4f^35d}{5}{L}{\circ}{7} - \eu{4f^36p}{5}{K}{}{6}  & ~~9 - 118 \\
5292.62$^*$ & 0.65 &\eu{4f^35d}{5}{K}{\circ}{6} - \eu{4f^36p}{5}{K}{}{6}  & ~10 - 118 \\
5322.77$^*$ & 0.48 &\eu{4f^35d}{5}{L}{\circ}{6} - \eu{4f^36p}{5}{K}{}{5}  & ~~6 - 110 \\
5681.88     & 1.16 &\eu{4f^35d}{5}{H}{\circ}{5} - \eu{4f^25d^2}{3}{G}{}{5}& ~26 - 134 \\
6017.80$^*$ & 1.11 &\eu{4f^35d}{5}{G}{\circ}{2} - \eu{4f^36p}{5}{H}{}{3}  & ~23 - 126 \\
6165.94$^*$ & 0.92 &\eu{4f^35d}{5}{I}{\circ}{4} - \eu{4f^36p}{5}{I}{}{4}  & ~15 - 116 \\
6656.83$^*$ & 1.82 &\eu{4f^35d}{3}{L}{\circ}{9} - \eu{4f^36p}{3}{K}{}{8}  & ~60 - 146 \\
\multicolumn{4}{l}{ \ \ \ion{Pr}{iii}} \\
4910.82 & 0.17 &\eu{4f^3}{4}{I}{\circ}{11/2} - \eu{4f^25d}{4}{H}{}{11/2} &296 - 322 \\
4929.12 & 0.36 &\eu{4f^3}{4}{I}{\circ}{13/2} - \eu{4f^25d}{4}{H}{}{13/2} &297 - 325 \\
5284.69 & 0.17 &\eu{4f^3}{4}{I}{\circ}{11/2} - \eu{4f^25d}{4}{H}{}{9/2}  &296 - 319 \\
5299.99 & 0.36 &\eu{4f^3}{4}{I}{\circ}{13/2} - \eu{4f^25d}{4}{H}{}{11/2} &297 - 322 \\
5844.41 & 1.24 &\eu{4f^3}{2}{H}{\circ}{9/2}  - \eu{4f^25d}{2}{G}{}{7/2}  &300 - 332 \\
5998.97 & 0.17 &\eu{4f^3}{4}{I}{\circ}{11/2} - \eu{4f^25d}{4}{G}{}{9/2}  &296 - 315 \\
6053.00 & 0.00 &\eu{4f^3}{4}{I}{\circ}{9/2} - \eu{4f^25d}{4}{G}{}{7/2}   &295 - 311 \\
6090.01 & 0.36 &\eu{4f^3}{4}{I}{\circ}{13/2} - \eu{4f^25d}{4}{H}{}{11/2} &297 - 318 \\
6160.23 & 0.17 &\eu{4f^3}{4}{I}{\circ}{11/2} - \eu{4f^25d}{4}{H}{}{9/2}  &296 - 314 \\
6195.62 & 0.00 &\eu{4f^3}{4}{I}{\circ}{9/2} - \eu{4f^25d}{4}{H}{}{7/2}   &295 - 310 \\
6500.04 & 1.72 &\eu{4f^3}{2}{G}{\circ}{7/2}  - \eu{4f^25d}{2}{F}{}{5/2}  &306 - 334 \\
6616.46 & 1.55 &\eu{4f^3}{2}{H}{\circ}{11/2}  - \eu{4f^25d}{4}{F}{}{9/2} &303 - 332 \\
6692.25 & 1.16 &\eu{4f^3}{4}{F}{\circ}{3/2} - \eu{4f^25d}{4}{F}{}{3/2}   &299 - 326 \\
6706.70 & 0.55 &\eu{4f^3}{4}{I}{\circ}{15/2} - \eu{4f^25d}{2}{I}{}{13/2} &298 - 318 \\
\noalign{\smallskip}\hline
\multicolumn{4}{l}{$ ^1$  Lines used only for solar abundance analysis.} \\
\multicolumn{4}{l}{$ ^2$  The level numbers in the model atom.} \\
\multicolumn{4}{l}{$ ^*$  HFS is taken into account.} \\
\end{tabular}
\end{table} %

The non-LTE effect on the line strength is determined by the
departures from LTE for the lower and upper levels of the
transition in the line formation layers. For the model 7250/4.3/0, the non-LTE effects are very small for the \ion{Pr}{ii}\, lines.
The obtained overpopulation of the \ion{Pr}{iii}\, levels leads to the
strengthening of the \ion{Pr}{iii}\, lines compared with the LTE case.
The theoretical non-LTE and LTE equivalent widths of the selected lines
and the non-LTE abundance corrections $\Delta_{\rm NLTE}$ =
$\eps{NLTE}$ -- $\eps{LTE}$ are presented in Table\,\ref{HD24712} (columns
6 -- 8).
Oscillator strengths used in these calculations are taken from Kurucz \& Bell (\cite{Kur94b}) for the \ion{Pr}{ii} lines and are computed in this paper for the lines of \ion{Pr}{iii}.

%
\begin{figure}
\begin{center}
\resizebox{88mm}{!}{\includegraphics{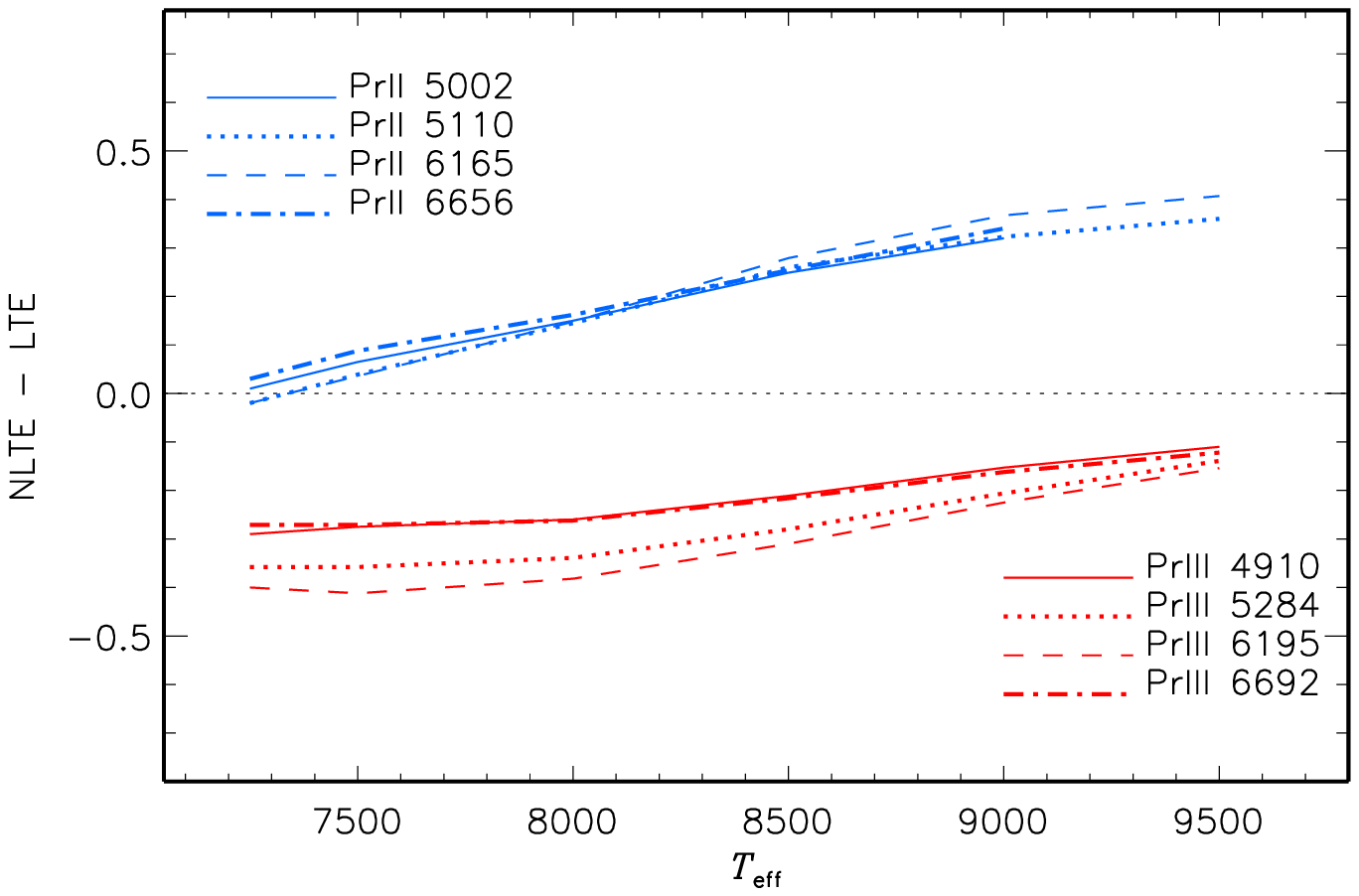}}
\caption[]{Non-LTE abundance corrections for the
\ion{Pr}{ii}\, and \ion{Pr}{iii}\, lines depending on $\Teff$. For
all models, $\logg$ = 4. The calculations were made with the Pr abundance [Pr/H] = 3.} \label{nlte_teff}
\end{center}
\end{figure}

As expected, the departures from LTE for the \ion{Pr}{ii}\, lines grow with
$\Teff$.
The larger $\Teff$, the stronger ultraviolet radiation is, thus resulting in
amplified overionization of \ion{Pr}{ii}. Figure\,\ref{nlte_teff} shows the calculated non-LTE abundance corrections for some lines of \ion{Pr}{ii}\, and \ion{Pr}{iii}\, depending on $\Teff$. Everywhere, [Pr/H] = 3 is adopted. For the \ion{Pr}{ii}\, lines, $\Delta_{\rm NLTE}$ are positive. They are small at $\Teff$ = 7250~K and 7500~K with $\Delta_{\rm NLTE} <$ 0.1~dex and grow to approximately 0.4~dex at $\Teff$ = 9500~K. In the hottest model, \ion{Pr}{ii} $\lambda$5002 and $\lambda$6656 have $W_\lambda \le$ 1~m\AA, and their $\Delta_{\rm NLTE}$ are not shown.  
In contrast, the departures from LTE for \ion{Pr}{iii}\, are weakened toward higher $\Teff$ because doubly-ionized praseodymium tends to represent the majority of the element and to preserve the TE total number density. The non-LTE abundance corrections are negative for the \ion{Pr}{iii}\, lines with $\Delta_{\rm NLTE} \simeq (-0.3) - (-0.4)$~dex at $\Teff$ = 7500~K and $\Delta_{\rm NLTE} \simeq -0.1$~dex at $\Teff$ =
9500~K. 


\begin{figure}
\resizebox{88mm}{!}{\includegraphics{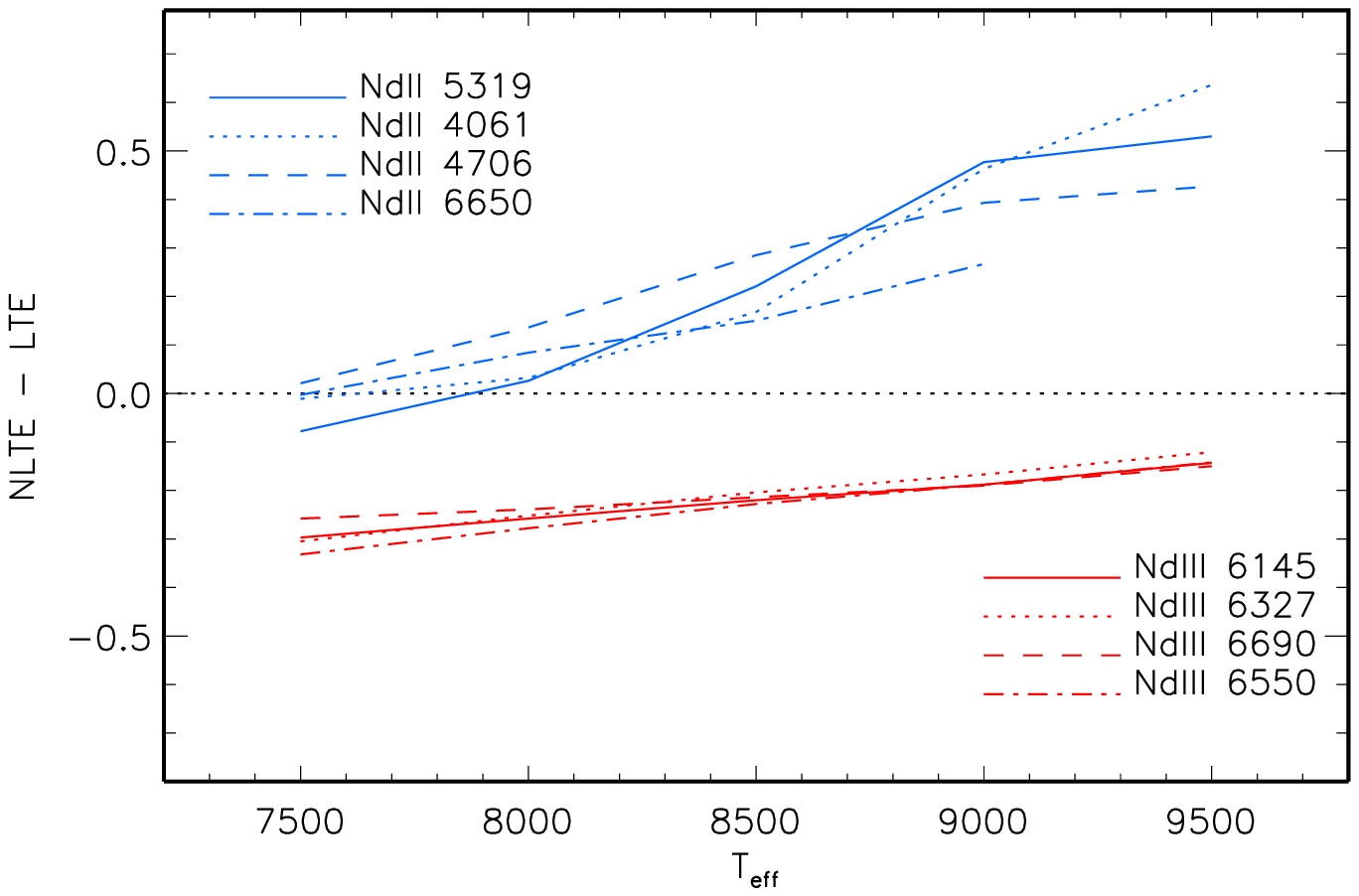}} 
\caption[]{Non-LTE abundance corrections for the \ion{Nd}{ii} and
\ion{Nd}{iii} lines depending on $\Teff$. For
all models, $\logg$ = 4. The calculations were made with the Nd abundance [Nd/H] = 2.5.}
\label{nlte_teff_nd}
\end{figure}

In this paper, we present also the revised non-LTE abundance corrections for
the \ion{Nd}{ii} and \ion{Nd}{iii}
lines (Fig.\,\ref{nlte_teff_nd}). Opacity package of the DETAIL code was
recently updated by the inclusion of the quasi-molecular Lyman $\alpha$
satellites following Castelli \& Kurucz (\cite{cast_kur2001}) implementation
of the Allard et~al. (\cite{allard98}) theory and by the use of the extended
line list based on not only measured but also predicted lines from Kurucz \& Bell 
(\cite{Kur94b}). 
The effect of the increased UV opacity below
1700\AA\, is seen only for the lines of \ion{Nd}{ii} at $\Teff \ge$ 8500~K.
At the lower temperatures, the revised $\Delta_{\rm NLTE}$ of the
\ion{Nd}{ii} lines agree within 0.01~dex to 0.03~dex with that obtained by
Mashonkina et~al. (\cite{mash_nd_ap}). For the \ion{Nd}{iii} lines, the
maximal difference between the revised and 2005's $\Delta_{\rm NLTE}$
constitutes 0.05~dex at $\Teff$ = 8500~K and is smaller at the
lower and higher temperatures. Thus, the Nd abundance distributions in the
atmospheres of HD\,24712 ($\Teff$ = 7250~K) and $\gamma$~Equ ($\Teff$ =
7700~K) found in our earlier paper do not need to be revised. For the
\ion{Nd}{ii} lines at $\Teff \ge$ 8500~K, we obtain now the smaller
departures from LTE due to decreased ionizing radiation. The difference in
$\Delta_{\rm NLTE}$ equals 0.1~dex to 0.2~dex for different lines at
different temperatures.

\subsection{The atmospheres with stratified distribution of praseodymium}

\begin{figure} 
\begin{center}
\resizebox{88mm}{!}{\includegraphics{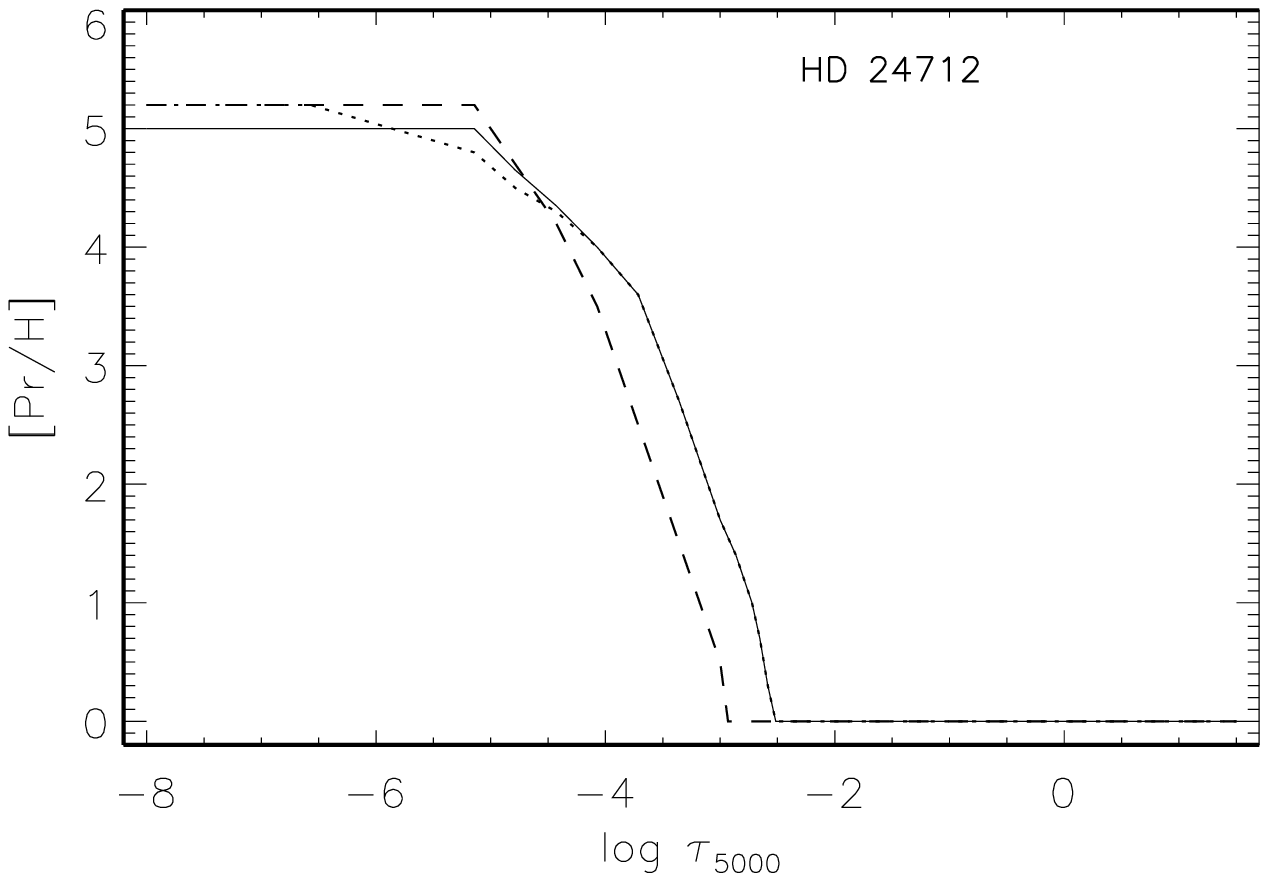}}

\vspace{-5mm} 
\resizebox{88mm}{!}{\includegraphics{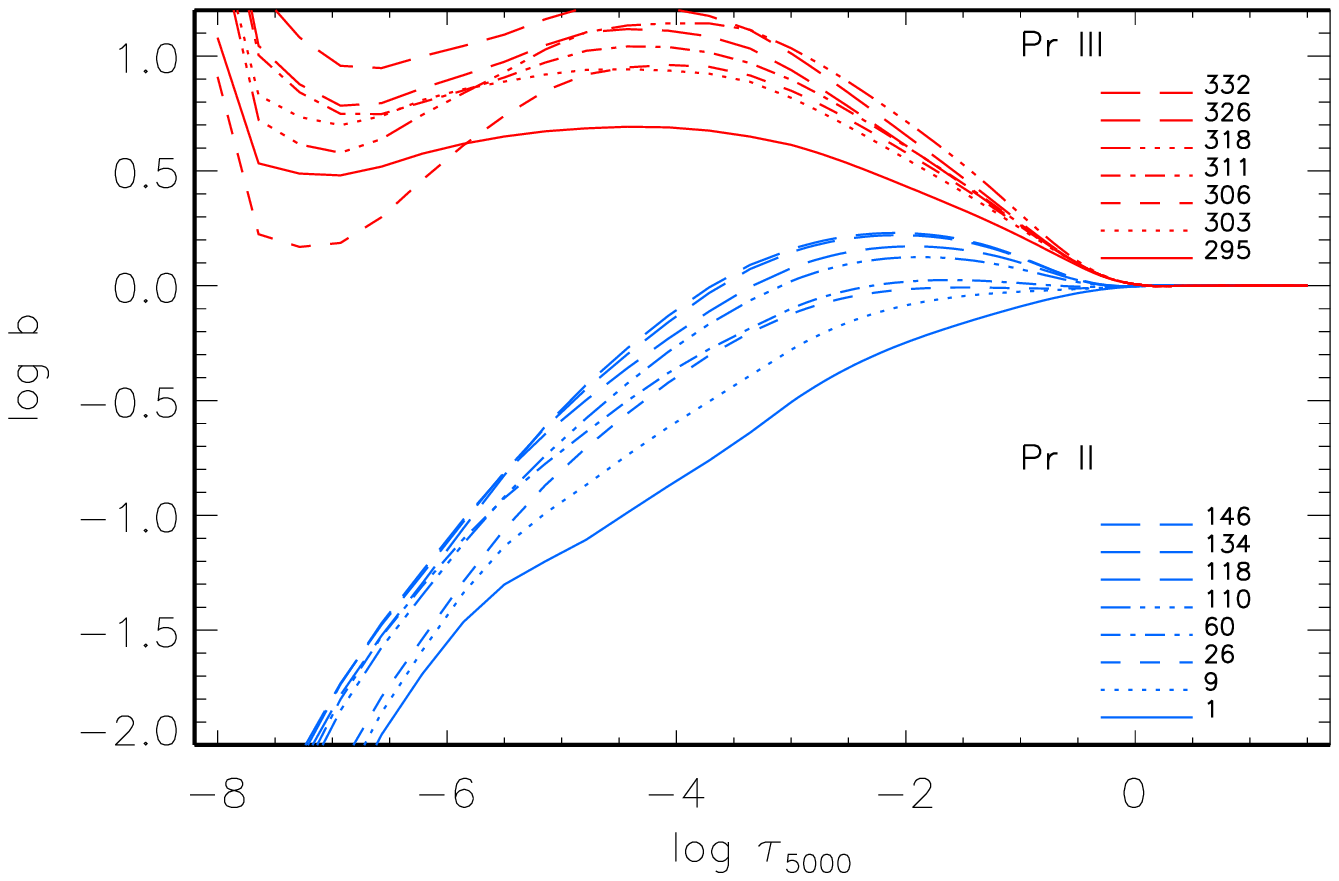}}
\caption[]{Top panel: the [Pr/H] ratio distributions found in the atmosphere of HD\,24712 (continuous line) and from the test calculations with 2~dex smaller photoionization cross-sections (dashed line) and with a variable effective collision strength for forbidden transitions (dotted line). 
Bottom panel: the departure coefficients $\log b$ for the selected levels
of \ion{Pr}{ii} and \ion{Pr}{iii} in the model 
7250/4.3/0 representing the atmosphere of HD\,24712 from
calculations with the praseodymium distribution shown in the
top panel by continuous curve. See text for more details.} \label{bf_layer}
\end{center}
\end{figure}

 In this section, the departures from LTE for \ion{Pr}{ii/iii} are investigated in the model 7250/4.3/0 with the stratified Pr 
abundance distribution determined in Sect.\,\ref{pr_hd24712} and shown by continuous curve in the top panel of Fig.\,\ref{bf_layer}.
The departure coefficients for the selected levels of \ion{Pr}{ii}\, and \ion{Pr}{iii}\, are shown in the bottom panel of the same figure. The
theoretical non-LTE and LTE equivalent widths together with non-LTE
abundance corrections are given in Table \ref{HD24712} (columns 9 -- 11).

Non-LTE calculations show the depletion of \ion{Pr}{ii}\, and enhanced number
density of \ion{Pr}{iii}\, in the line formation layers similar to the
case of homogeneous Pr abundance distribution. However, three important
distinctions in the behavior of departure coefficients can be seen.
\begin{itemize}
\item The magnitude of the depletion of \ion{Pr}{ii}\, is much larger in the
stratified atmosphere, in particular, above $\log\tau_{5000} = -4$ where
the lines are formed.
\item In the stratified atmosphere, for every pair of the \ion{Pr}{ii}\,
levels with $i < j$, an inequality $b_i < b_j$ is valid.
\item Above $\log\tau_{5000}$ = --4 in the stratified atmosphere, the
excited levels of \ion{Pr}{iii}\, are decoupled to the ground state to the
more extent compared to the case of the homogeneous Pr abundance distribution.
\end{itemize}

In the stratified atmosphere, the Pr lines are formed in the uppermost atmospheric layers, above $\log\tau_{5000} = -4$, where the departures from LTE are large.
Non-LTE leads to weakening the \ion{Pr}{ii}\, lines due to the decreased number of absorbers ($b_i < 1$) and due to the line source function $S_{ij} \simeq b_j/b_i B_\nu(T_e)$ 
exceeding the Planck function ($b_i < b_j$).  The
non-LTE abundance corrections for various \ion{Pr}{ii}\, lines are positive and may reach +1.2~dex. The \ion{Pr}{iii}\, lines are strengthened compared with
the LTE case, and $\Delta_{\rm NLTE}$ may reach --0.7~dex. Since the non-LTE effects have the opposite sign for the \ion{Pr}{ii}\, and \ion{Pr}{iii}\, lines,
they are important for the comparison of Pr abundances derived from these lines.

\section{Solar abundance of praseodymium}\label{sun}

We first apply the non-LTE calculations to the Pr abundance analysis for the solar atmosphere. The earlier determinations were performed by 
Bi\'emont et~al. (\cite{B79}) based on the nine lines of \ion{Pr}{ii} in the 3990 -- 5330~\AA\ spectral region and by Ivarsson et~al. (\cite{ILW2003}) 
based on the three lines of \ion{Pr}{ii}. The obtained mean abundances of the praseodymium are surprisingly different, $\log\rm {(Pr/H)}_\odot=-11.29\pm0.08$ 
and $\log\rm {(Pr/H)}_\odot=-11.60\pm0.1$, respectively. The most recent study is based on the modern accurate laboratory measurements for oscillator 
strengths of Ivarsson et~al. (\cite{ILW}). Ivarsson et~al. (\cite{ILW2003}) also test the influence of 3D effects on the determination of the Pr abundance 
and draw a preliminary conclusion that the differences in the equivalent widths of the \ion{Pr}{ii} lines between the 3D and 1D cases are insignificant.

Each of the lines used by our predecessors was checked for blending using the NSO solar flux spectrum (Kurucz et~al. 1984) observed with the spectral 
resolving power $R \simeq 340 000$ at wavelengths between 4000\AA\ and 4700\AA\ and with $R = 520 000$ at longer wavelengths. We find four lines suitable for 
precise spectral fitting. They are listed in 
Table~\ref{solar}. 
Spectral region around each investigated line was synthesised with the {\sc SIU} code (Reetz \cite{Reetz}). For given atomic level, 
{\sc SIU} computes the non-LTE population as the production of the LTE occupation number and the corresponding departure coefficient. 
In LTE calculations, the revised \ion{Pr}{ii} partition function is applied. 
Calculations of the \ion{Pr}{ii} lines were made with the three different sets of transition probabilities taken from Kurucz \& Bell (\cite{Kur94b}), Ivarsson et~al. (\cite{ILW}), and Li et~al. (\cite{LCH}). 
Atomic parameters for other atomic lines in the synthesised regions are taken from the latest release of {\sc VALD} (Kupka et~al. \cite{vald}). For molecular lines, we apply the data compiled by Kurucz (\cite{Kur94a}).
We use the theoretical {\sc ATLAS9} model atmosphere of the Sun (5777/4.44/0, Heiter et~al. \cite{sun_cgm}) with convection treated according to Canuto et~al. (\cite{cgm}) and the semi-empirical model of 
Holweger \& M\"uller (\cite{HM}, HM). A microturbulence velocity \Vmic = 0.9\,\kms\ is adopted. Our synthetic flux profiles are convolved with a profile that combines a rotational broadening of 1.8~\kms, broadening by macroturbulence with a 
radial-tangential profile of \Vmac = 3.6\,\kms, and instrumental broadening with a Gaussian profile corresponding to the spectral resolution. A macroturbulence velocity was allowed to vary within 0.2\,\kms\ to achieve the best fit to the observed line shape. 

\begin{figure*} 
\begin{center}
\resizebox{140mm}{!}{\includegraphics{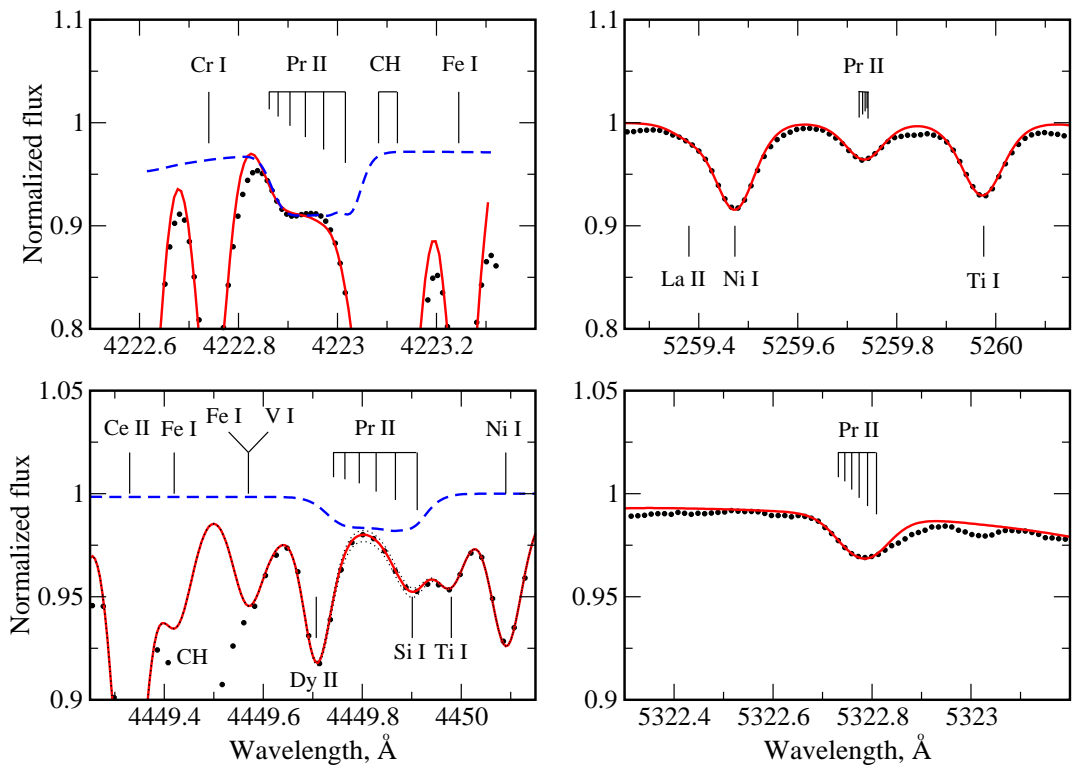}}
\caption[]{Synthetic spectra (continuous curve) and pure \ion{Pr}{ii} line profiles (dashed curve, for $\lambda$4223~\AA\, and $\lambda$4449~\AA\
lines only) calculated with the {\sc ATLAS9} solar model atmosphere in comparison with the observed NSO solar flux spectrum 
(Kurucz et~al. 1984, bold dots). For the 4449~\AA\ blend, dotted lines show the synthetic spectra computed with $\pm$0.05~dex 
change in the adopted Pr abundance. The \ion{Pr}{ii} lines are treated based on non-LTE line formation.}
\label{pr_sun}
\end{center}
\end{figure*}

The best fits to the observed solar line profiles are shown in Fig.~\ref{pr_sun}. The results from calculations with the {\sc ATLAS9} solar model atmosphere 
are presented in Table~\ref{solar}. 
 Below we give brief notes on the individual lines.

\underline{\ion{Pr}{ii} 4222.93~\AA.} The continuum flux at 4222.93~\AA\ is influenced by the far wing of the \ion{Ca}{i} resonance line at 
4226.7~\AA. The \ion{Ca}{i} line is treated using the measured oscillator strength from Smith \& Gallagher (\cite{ca4226}) and the van der Waals broadening 
parameters based on the advanced perturbation theory of Anstee \& O'Mara (\cite{omara_sp}). The \ion{Pr}{ii} 4222.93\AA\ line is partially ovelapped
with the blue wing of the spectral feature produced by two CH lines at 4223.091\AA\ and 4223.113\AA. A half-width of this spectral feature is only slightly affected by the \ion{Pr}{ii} line and is fitted reasonably well if we reduce oscillator strengths for both CH lines by 0.3~dex compared to those 
given by Kurucz (\cite{Kur94a}). 
Small changes in wavelengths (no more than 0.014~\AA) and oscillator strengths (no more than 0.27~dex) were introduced to fit the \ion{Cr}{i} 4222.740~\AA\ and 
\ion{Fe}{i} 4223.237~\AA\ lines just to make better fit.

\underline{\ion{Pr}{ii} 5259.7~\AA.} This line is well isolated and very well fitted by spectrum synthesis. The \ion{Ni}{i} 5259.466~\AA\ and 
\ion{Ti}{i} 5259.973\AA\ lines shown in Fig.~\ref{pr_sun} do not affect the Pr abundance determination. For the \ion{Ni}{i} line, changes in wavelength 
 by 0.011~\AA\ and in oscillator strength by $-0.4$~dex were made to fit the line profile, while a 0.005~\AA\ wavelength change is required for \ion{Ti}{i} line.

\underline{\ion{Pr}{ii} 5322.8~\AA} is well isolated, too, but the continuum flux 
is influenced by the far wing of the \ion{Fe}{i} 5324.179~\AA\ line. With the best atomic parameters of the \ion{Fe}{i} line, $\log gf = -0.103$ (Bard et~al. \cite{BKK}) 
and $\log \gamma_6/N_H$ = --7.035 at $T_e$ = 10000~K (increased by 0.2~dex compared with the Anstee \& O'Mara's \cite{omara_sp} value), the predicted synthetic spectrum is still
0.2\%\ higher than the observed spectrum around 5323~\AA. The local continuum level was, therefore, adjusted to fit the blue wing of \ion{Fe}{i} 5324~\AA. A variation 
of 0.2\%\ in the continuum flux produces a 0.05~dex change in the Pr abundance derived from \ion{Pr}{ii} 5322.8~\AA.

\begin{table*}
\begin{center}
\caption{Solar praseodymium abundance from calculations with the {\sc ATLAS9} solar model atmosphere.
} \label{solar}
\setlength{\tabcolsep}{4.5mm} 
\begin{tabular}{c|ccc|ccc|c}
\hline\noalign{\smallskip}
 $\lambda$ [\AA] & \multicolumn{3}{c|}{$\log gf$}&\multicolumn{3}{c|}{$\log\rm {(Pr/H)_{LTE}}$}&$\Delta_{\rm NLTE}$  \\
\cline{2-7}
         & KB$^1$& ILW$^2$&  LCH$^3$&  KB      &  ILW     &   LCH   &       \\   
\hline
4222.93 &~0.13   &~0.27   &~0.24   &$-$11.14 &$-$11.28  &$-$11.25&$+$0.03 \\
4449.83 &$-$0.32 &$-$0.17&$-$0.26 &$-$11.07 &$-$11.22  &$-$11.13&$+$0.02 \\
5259.73 &~0.08   &~0.11   &~0.07   &$-$11.11 &$-$11.14  &$-$11.10&$+$0.04 \\
5322.77 &$-$0.32 &$-$0.32&$-$0.12 &$-$11.04 &$-$11.04  &$-$11.24&$+$0.02 \\
\hline
 &\multicolumn{3}{c|}{~~~~LTE-average}&$-11.09\pm0.04$&$-11.17\pm0.10$&$-11.18\pm0.08$ &\\
 &\multicolumn{3}{c|}{non-LTE-average}&$-11.06\pm0.04$&$-11.14\pm0.10$&$-11.15\pm0.08$ &\\
\hline
\multicolumn{3}{l}{~~$^1$ Kurucz \& Bell (\cite{Kur94b}, KB)} & 
\multicolumn{2}{l}{~~$^2$ Ivarsson et~al. (\cite{ILW}, ILW)} &
\multicolumn{3}{l}{~~$^3$ Li et~al. (\cite{LCH}, LCH)} \\
\end{tabular}
\end{center}
\end{table*}
    
\underline{\ion{Pr}{ii} 4449.83~\AA} is located between the \ion{Dy}{ii} 4449.70~\AA\ and \ion{Si}{i} 4449.90~\AA\ lines. The blue wing and the core of the 
\ion{Dy}{ii} line are slightly affected by the adjacent lines and are well fitted using a center line wavelength of 4449.707~\AA, accurate experimental oscillator strength, $\log gf = -1.03$ 
(Wickliffe et~al. \cite{WLN}), and the solar Dy abundance $\eps{Dy,\odot}$ = 1.14 (Lodders \cite{Lod}). 
The \ion{Si}{i} line is predicted, and its oscillator strength has to be reduced down to $\log gf$ = --3.33 to fit the line core that is only slightly affected 
by the HFS components of the \ion{Pr}{ii} line. In Fig.~\ref{pr_sun}, we show the best fit of the 4449~\AA\ blend and also the synthetic spectra computed with 
$\pm$0.05~dex change in adopted Pr abundance. The uncertainty of the element abundance derived from \ion{Pr}{ii} 4449 is, obviously, does not exceed 0.05~dex. 


In the solar atmosphere, \ion{Pr}{ii}\, represents the majority of the element, and its ground state and the low excitation levels which are the lower levels of the 
investigated transitions preserve the TE number density. Our non-LTE calculations show that photon pumping produces enhanced excitation of the upper levels 
resulting in weakening the lines of interest compared to the LTE case. We find that the statistical equilibrium of \ion{Pr}{ii} in the solar atmosphere is sensitive to a variation of collision excitation rates which include, for the Sun, interactions with not only electrons but also neutral hydrogen atoms. For hydrogenic collisions, we use the formula of Steenbock \& Holweger (\cite{hyd}) for allowed transitions and follow Takeda (\cite{Takeda1994}) for forbidden transitions. Both theoretical approximations provide only an order of magnitude estimate. 
Empirical constraining the efficiency of hydrogenic collisions in the SE of various atoms that is represented by a scaling factor
\kH\ applied to the mentioned formula gives a variety of estimates between \kH\ = 0.002 and \kH\ = 1 (for review, see Mashonkina \cite{ml_eas}).  When pure electronic 
collisions are taken into
account (\kH\ = 0), the non-LTE abundance corrections constitute +0.03~dex to +0.08~dex for different lines. The non-LTE effects become negligible 
($\Delta_{\rm NLTE} \le$ 0.01~dex) when both types of collisions are included and \kH\ = 1. We adopt an intermediate value, \kH\ = 0.1. 
The corresponding non-LTE corrections are shown in the last column of Table~\ref{solar}. 

The average LTE and non-LTE praseodymium abundances from calculations with the {\sc ATLAS9} solar model atmosphere and their standard deviations 
are presented in two bottom strings of Table~\ref{solar}. The Pr abundance determined with the HM model is larger, by 0.05~dex, on average. It is worth noting that the statistical error of our abundance determination estimated from the line-to-line scatter corresponds, in fact, to the quoted errors of $\log gf$ measurements which varies between  0.03~dex and 0.07~dex for the lines in Table~\ref{solar}. 
The three solutions found with $\log gf$ from different sources are consistent within the error bars. However, the preference should be given to the 
ILW and LCH abundances based on the modern experimental transition probabilities.  
The average ILW and LCH abundances agree within 0.01~dex.
For each set of atomic parameters and for both solar model atmospheres, the derived Pr abundance is 0.09~dex -- 0.23~dex larger than that obtained by 
Bi\'emont et~al. (\cite{B79}) and 0.40~dex -- 0.54~dex larger compared to that of Ivarsson et~al. (\cite{ILW2003}). Different studies give the meteoritic CI Chondrites abundance  
$\log\rm {(Pr/H)}_{met}$ = --11.20$\pm0.04$ 
(Grevesse et~al. \cite{met96}; Palme \& Jones \cite{Palme2005}),  --11.22$\pm0.03$ (Lodders \cite{Lod}) or --11.25$\pm0.03$ (Asplund et~al. \cite{met05}).
Our best solar Pr abundance for the {\sc ATLAS9}/HM model, $\log\rm {(Pr/H)}_\odot=-11.15\pm0.08$ / --11.10$\pm$0.08, exceeds the meteoritic values by 0.05~dex to 0.10~dex. 

In solar abundance determinations, we use the partition function calculated by its definition and based on an extensive set of the predicted energy levels of \ion{Pr}{ii}, 
$U_{calc}$(\ion{Pr}{ii}). As can be seen from Table\,\ref{pf}, $U_{calc}$(\ion{Pr}{ii}) is about 30\%\ larger than that for the experimental energy 
levels only, $U_{exp}$(\ion{Pr}{ii}), at temperatures of the line formation layers in the solar atmosphere. 

With $U_{exp}$(\ion{Pr}{ii}),  
the inferred solar Pr abundance is 
$\log\rm {(Pr/H)_\odot}$ = --11.22 and --11.17, for the {\sc ATLAS9} and the Holweger \& M\"uller (\cite{HM}) models, respectively. Thus, accurate solar 
Pr abundance depends on the completeness of 
known \ion{Pr}{ii} levels and on the accuracy of their energies.
It seems that the current theoretical energy calculations give us the upper limit for the solar Pr abundance.

\section{Praseodymium in the roAp star HD\,24712} \label{obs}

We study the Pr abundance in the atmosphere of the roAp star HD\,24712 using its average spectrum obtained during spectroscopic
monitoring on November 11/12, 2004 with the UVES 
spectrograph at the 8.2-m telescope, UT2 (Kueyen), of the VLT on Paranal, Chile (programme 274.D-5011). Observations and data reduction
are described by Ryabchikova et~al. (\cite{Ryabchikova2007a}). The spectrum has been obtained near the magnetic maximum, that coincides
roughly with the REE spectral line intensity maximum. The surface magnetic field 
measured at the magnetic maximum is \bs\,= 3.1 -- 3.3~kG (Ryabchikova et~al. 
\cite{Ryabchikova2007a}). 

\subsection{The problem of the Pr abundance in HD\,24712}\label{star_uniform}

We start from element abundance determination using the lines of \ion{Pr}{ii} and \ion{Pr}{iii} listed in Table\,\ref{HD24712} and assuming the uniform
distribution of Pr in the atmosphere of HD\,24712. The choice of the spectral lines was determined mainly by the possibility of pulsation measurements in the HD\,24712 spectrum. 
Magnetic spectrum synthesis is performed with the help of {\sc SYNTHMAG} code 
under the LTE assumption using the
model atmosphere 7250/4.3/0. We apply most recent experimental transition probabilities by Ivarsson et~al. (\cite{ILW}) and Li et~al. (\cite{LCH}) if available and oscillator strengths from  Kurucz \& Bell (\cite{Kur94b}) for the remaining \ion{Pr}{ii} lines. 
The derived LTE abundance is $\log\rm {(Pr/H)}=-9.4\pm0.2$ from 10 \ion{Pr}{ii} lines 
and by 2.1~dex larger, $\log\rm {(Pr/H)} = -7.3\pm0.3$, from 14 \ion{Pr}{iii} lines. Such a large discrepancy between 
the rare-earth LTE abundances derived from two ionization stages is typically observed in the roAp and cool Ap stars (the REE anomaly, 
see Ryabchikova et~al. \cite{Ryabchikova2004}). 

Our non-LTE code does not include the magnetic field in statistical equilibrium calculations. 
 The presence of a magnetic field should not cause significant changes in the derived level populations, for moderate field strengths. 
As was shown in Sect.\,\ref{departures}, departures from LTE for the level populations are mainly caused by UV continua and by strong UV transitions. 
The change in emergent continuum flux around 2000\AA\ due to the influence of the magnetic field on atmospheric structure does not exceed 
7\%, for field strengths up to 10~kG (Kochukhov et~al. \cite{Ketal05}). The critical strong UV transitions have small Zeeman splittings, since 
the splittings scale with the wavelength. Further abundance determinations are, therefore, performed in the field-free approximation.

 With fixed level populations, the magnetic field can affect an abundance analysis through Zeeman splitting of spectral lines. 
Therefore, we investigate a possibility to replace magnetic broadening effects 
by other broadening mechanisms, for instance, by a microturbulence. 
Different spectral lines used in this study have different Zeeman patterns. The \ion{Pr}{ii} lines are weak and HFS-affected. We adopt a common 
value, \Vmic\,= 1\,\kms, in their analysis. 
For each line of \ion{Pr}{iii}, a pseudo-microturbulence velocity is calculated as follows. We first fit the line profile with the {\sc SYNTHMAG} 
code to derive the Pr abundance in LTE approximation. Then, calculations are made with the {\sc SYNTH3} code (Kochukhov \cite{synth3}) that ignores 
the existence of magnetic field. 
The element abundance is fixed and \Vmic\, varies until the theoretical equivalent width reaches the observed one.  The individual values of \Vmic\ 
are given in Table\,\ref{HD24712}. Applying this procedure we find only minor changes in the element abundances, by 0.14~dex and 0.16~dex from the \ion{Pr}{ii} and 
\ion{Pr}{iii} lines, respectively, compared to those determined from magnetic spectrum synthesis. It is worth noting, that the abundance difference 
between two ionization stages remains to be almost the same.

Next, we check whether the departures from LTE can solve the problem of the praseodymium abundance in HD\,24712 and find that non-LTE tends to reduce the difference in LTE abundances 
but fails to remove it completely. 
The non-LTE abundance is $\log\rm {(Pr/H)}=-9.22\pm0.19$ from the \ion{Pr}{ii} lines, and a 1.90~dex larger value is determined from the lines of 
\ion{Pr}{iii}. 
 Thus, we cannot obtain consistent abundances of the praseodymium from the lines of two ionization stages assuming the uniform element distribution in the atmosphere 
of HD\,24712, either considering non-LTE line formation, or taking into account the splitting of the spectral lines in magnetic field.
We suggest, therefore, that similarly to the neodymium (Mashonkina et~al. \cite{mash_nd_ap}) the praseodymium is distributed non-uniformly in the atmosphere 
of HD\,24712, although the properties of the Nd and Pr distributions may be different.

\subsection{Vertical distribution of Pr in the atmosphere}\label{pr_hd24712}

Using the simplified step-function approximation as an initial guess for the Pr abundance profile, we modify it by the trial-and-error method based on non-LTE line formation trying to fit the observed equivalent widths of both \ion{Pr}{ii} and \ion{Pr}{iii} lines in the spectrum of HD\,24712. Due to certain limitations of the present non-LTE calculations, mainly due to ignoring Zeeman splitting, we cannot use the observed
line profiles in stratification analysis as this was done, for instance, in LTE analysis for Ca (Ryabchikova et.~al \cite{Ca-IS}).
 The final solution is shown in the top panel of Fig.\,\ref{bf_layer}. The praseodymium is concentrated mostly above $\log \tau_{5000} = -4$, and the 
required Pr abundance in the layer is [Pr/H] $\ge$ 4.
Table\,\ref{HD24712} (column 10) presents the corresponding non-LTE equivalent widths of the \ion{Pr}{ii}\ and \ion{Pr}{iii}\ lines. For comparison, we give also the LTE equivalent widths, calculated with the obtained Pr distribution, and the non-LTE abundance corrections.
It can be seen that net non-LTE correction to the Pr abundance $\Delta_{\rm NLTE}(\ion{Pr}{ii})-\Delta_{\rm NLTE}(\ion{Pr}{iii})$ is at the level of 1.8~dex, 
for the stratified atmosphere. In order to make the LTE element abundances from the lines of \ion{Pr}{ii} and \ion{Pr}{iii} consistent, the layer of enhanced 
praseodymium has to be located above the uppermost depth point in our model, $\log \tau_{5000} = -8$ (!). 
This means that any stratification analysis for the praseodymium in roAp stars has to be performed based on non-LTE line formation. It is worth noting that a 
location of the enriched praseodymium layer turns out to be very similar to that found for the neodymium in our earlier study.

\begin{figure}
\begin{center}
\includegraphics[width=88mm]{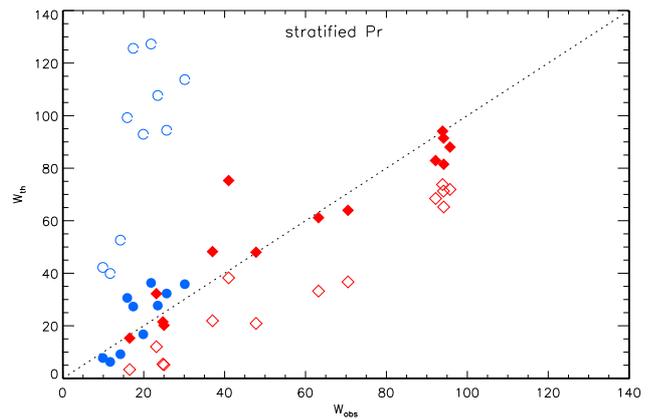}
\caption[]{Observed equivalent widths of the \ion{Pr}{ii}\,
(circles) and \ion{Pr}{iii}\, (diamonds) lines in HD\,24712
compared with the theoretical non-LTE (filled symbols) and LTE
(open symbols) equivalent widths for 
the stratified Pr distribution
shown in the top panel of Fig.~\ref{bf_layer} by continuous line.}
\label{obs_theory}
\end{center}
\end{figure}

\begin{figure*} 
\begin{center}
\resizebox{140mm}{!}{\includegraphics{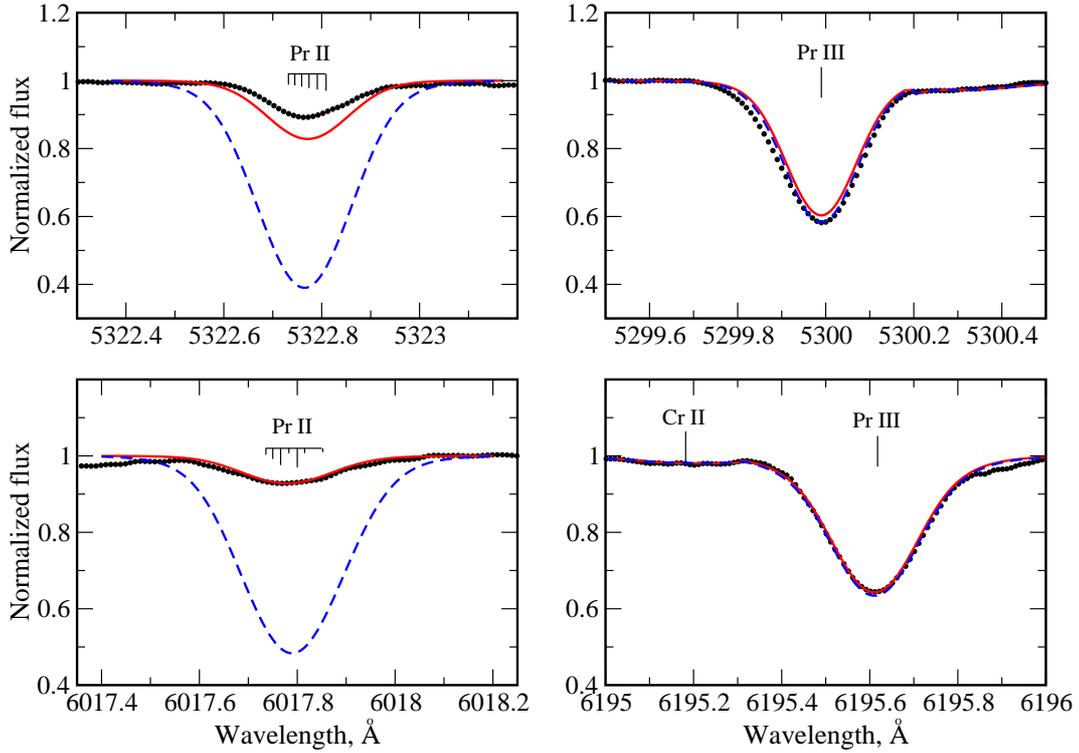}}
\caption[]{A comparison between the observed (bold dots) and non-LTE spectra for the selected \ion{Pr}{ii} and \ion{Pr}{iii}
lines in HD~24712. The non-LTE profiles from calculations with the derived Pr distribution are shown by continuous curves. Dashed curves correspond to the non-LTE profiles from calculations with log(Pr/H) = --7.32 everywhere in the atmosphere.}
\label{hd24712_lines}
\end{center}
\end{figure*}

A quality of fitting the observed praseodymium lines in HD\,24712 with the derived Pr stratification profile is illustrated in Fig.\,\ref{obs_theory} and Fig.\,\ref{hd24712_lines}. 
The synthetic flux profiles are convolved with the instrumental profile (spectral resolution R = 80000), projected rotational velocity \vs\ = 5.6~\kms\ and 
additional broadening parameter equivalent to radial-tangential profile of \Vmac = 4\,\kms\ for the \ion{Pr}{ii} and  6\,\kms\ for the \ion{Pr}{iii} lines.
One cannot expect to get the same quality fits for the magnetic star with a complex chemically stratified atmosphere as those for the Sun.
Nevertheless, our non-LTE stratification analysis allows to reproduce reasonably well the observed \ion{Pr}{ii} and \ion{Pr}{iii} lines in HD~24712, 
contrary to the case of the uniform distribution of Pr. In the latter case, we can fit the \ion{Pr}{iii} 5300\AA\ and 6195\AA\ profiles (dashed curves in Fig.\,\ref{hd24712_lines}) with the Pr abundance determined from non-LTE analysis of the \ion{Pr}{iii} lines only, log(Pr/H) = --7.32, 
but calculate extremely strong lines of \ion{Pr}{ii} compared to the observations. 

\begin{table*}
\begin{center}
\caption{Equivalent widths (in m\AA) of the \ion{Pr}{ii} and \ion{Pr}{iii} lines observed in
HD\,24712 (column 4) and calculated using non-LTE and LTE approach for the model 7250/4.3/0 with the
homogeneous (columns 6 - 8) and stratified (columns 9 - 11) element distributions.} 
\label{HD24712}
\begin{tabular}{lcrrl|rrr|rrr|c|c}
\hline\noalign{\smallskip}
  &  &  &  & &\multicolumn{3}{c|}{[Pr/H] = 3} &\multicolumn{5}{c}{Stratified
distribution$^*$} \\
\cline{6-13}
$\lambda$, & E$_{low}$, & $\log gf$ & W$_{obs}$ & \Vmic, &  & &  & &  & &
\multicolumn{1}{c|}{$\log \tau_{5000}^{l}$}&\multicolumn{1}{c}{$\log
\tau_{5000}^{l+c}$} \\
 \AA       & eV         &           & m\AA & \kms&\scriptsize{W$_{\rm LTE}$} &
\scriptsize{W$_{\rm NLTE}$} & \scriptsize{$\Delta_{\rm NLTE}$} &
\scriptsize{W$_{\rm LTE}$} & \scriptsize{W$_{\rm NLTE}$} &
\scriptsize{$\Delta_{\rm NLTE}$} & \scriptsize{NLTE} & \scriptsize{NLTE}\\
\hline\noalign{\smallskip}
 1 & 2 & \multicolumn{1}{c}{3} & \multicolumn{1}{c}{4} &\multicolumn{1}{c|}{5} &
\multicolumn{1}{c}{6} &
\multicolumn{1}{c}{7} & \multicolumn{1}{c|}{8} & \multicolumn{1}{c}{9} &
\multicolumn{1}{c}{10} &
\multicolumn{1}{c|}{11} & 12 & 13 \\
\hline\noalign{\smallskip}
\multicolumn{13}{l}{~~\ion{Pr}{ii}} \\
5002.44 & 0.80 & -0.87$^1$ & 12 & 1.0  & 31  &  30 &  0.02 & 40 &  6 & 1.12 & -4.50 &  -0.65 \\  
5110.76 & 1.15 &  0.32$^2$ & 30 & 1.0  & 95  &  96 & -0.02 &114 & 36 & 1.11 & -4.57 &  -1.86 \\  
5129.54 & 0.65 & -0.13$^1$ & 26 & 1.0  & 77  &  78 & -0.03 & 94 & 32 & 1.17 & -4.68 &  -2.08 \\  
5135.14 & 0.95 &  0.01$^3$ & 16 & 1.0  & 80  &  80 & -0.01 & 99 & 31 & 1.07 & -4.59 &  -1.47 \\  
5292.62 & 0.65 & -0.26$^3$ & 23 & 1.0  & 84  &  84 &  0.00 &108 & 28 & 1.19 & -4.60 &  -1.64 \\  
5322.77 & 0.48 & -0.32$^3$ & 22 & 1.0  & 99  & 100 & -0.01 &127 & 36 & 1.18 & -4.64 &  -1.44 \\  
5681.88 & 1.16 & -0.60$^2$ & 10 & 1.0  & 34  &  33 &  0.02 & 42 &  8 & 1.02 & -4.49 &  -0.75 \\  
6017.80 & 1.11 & -0.26$^2$ & 20 & 1.0  & 69  &  70 & -0.02 & 93 & 17 & 1.10 & -4.48 &  -1.03 \\  
6165.94 & 0.92 & -0.20$^2$ & 17 & 1.0  & 92  &  93 & -0.02 &126 & 27 & 1.13 & -4.55 &  -1.38 \\  
6656.83 & 1.82 &  0.08$^2$ & 14 & 1.0  & 41  &  39 &  0.03 & 53 &  9 & 1.06 & -4.47 &  -1.19 \\  
\multicolumn{13}{l}{~~\ion{Pr}{iii}}\\
4910.82 & 0.17 & -1.95$^4$ & 37 & 1.7  & 10  &  17 & -0.29 & 22 & 48 &-0.51 & -5.20 & -2.65 \\  
4929.12 & 0.36 & -2.07 & 23 & 1.0  &  6  &  11 & -0.27 & 12 & 32 &-0.61 & -5.17 & -2.06 \\  
5284.69 & 0.17 & -0.77 & 92 & 1.0  & 48  &  60 & -0.36 & 68 & 83 &-0.48 & -6.31 & -5.74 \\  
5299.99 & 0.36 & -0.72 & 94 & 1.05 & 44  &  58 & -0.36 & 65 & 81 &-0.48 & -6.24 & -5.66 \\  
5844.41 & 1.24 & -1.01 & 48 & 1.0  & 13  &  20 & -0.24 & 21 & 48 &-0.62 & -5.45 & -4.02 \\  
5998.97 & 0.17 & -1.87 & 63 & 1.55 & 14  &  24 & -0.29 & 33 & 61 &-0.45 & -5.47 & -3.81 \\  
6053.00 & 0.00 & -1.98 & 70 & 1.55 & 14  &  24 & -0.29 & 37 & 64 &-0.42 & -5.44 & -3.92 \\  
6090.01 & 0.36 & -0.87 & 94 & 1.1  & 46  &  61 & -0.36 & 71 & 91 &-0.53 & -6.19 & -5.58 \\  
6160.23 & 0.17 & -1.02 & 96 & 1.1  & 43  &  59 & -0.37 & 72 & 88 &-0.43 & -6.23 & -5.69 \\  
6195.62 & 0.00 & -1.07 & 94 & 1.0  & 46  &  63 & -0.40 & 74 & 94 &-0.57 & -6.19 & -5.63 \\  
6500.04 & 1.72 & -1.26 & 25 & 1.0  &  3  &   5 & -0.26 &  5 & 22 &-0.70 & -5.11 & -2.13 \\  
6616.46 & 1.55 & -1.50 & 25 & 1.0  &  3  &   5 & -0.25 &  5 & 20 &-0.69 & -5.07 & -1.92 \\  
6692.25 & 1.16 & -2.11 & 17 & 1.0  &  2  &   3 & -0.27 &  3 & 15 &-0.73 & -5.02 & -1.14 \\  
6706.70 & 0.55 & -1.49 & 41 & 1.5  & 18  &  30 & -0.31 & 38 & 75 &-0.58 & -5.49 & -4.22 \\  
\noalign{\smallskip}\hline
\multicolumn{13}{l}{$ ^1$ Li et~al. (\cite{LCH}); ~~~~$ ^2$ Kurucz \& Bell (\cite{Kur94b}); 
~~~$ ^3$ Ivarsson et~al. (\cite{ILW}); ~~~$ ^4$ this study, for all \ion{Pr}{iii} lines,} \\
\multicolumn{13}{l}{$ ^*$ It is shown in the bottom panel of Fig. \ref{bf_layer} by continuous curve.}
\end{tabular}
\end{center}
\end{table*} %

\begin{figure}
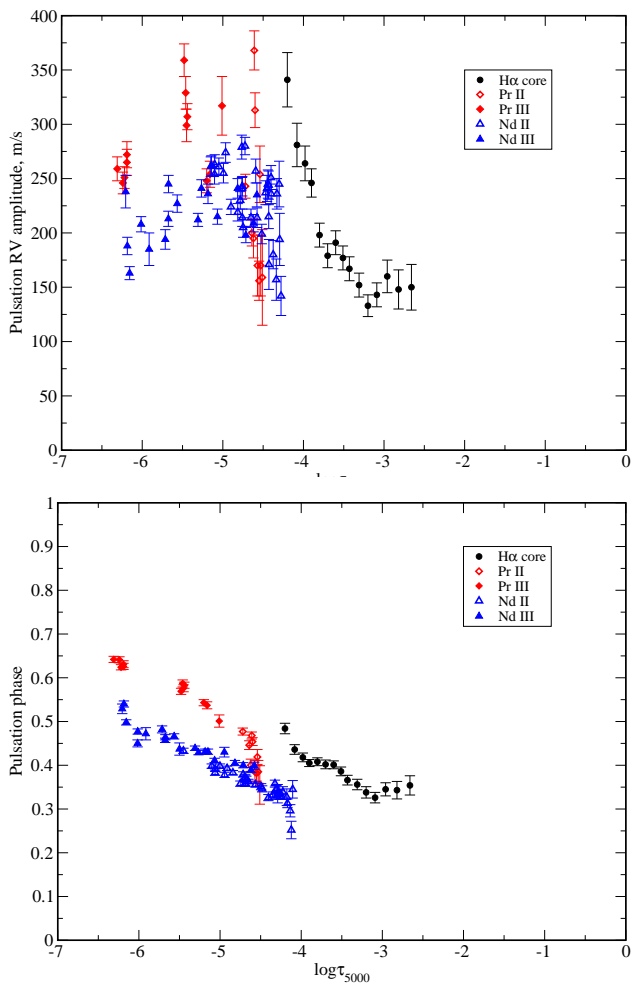

\begin{center}
\includegraphics[width=82mm]{UVES1_vr-tau_os.eps}\hspace{0.5cm}\includegraphics[width=82mm]{UVES1_ph-tau_os.eps}
\caption[]{Pulsation radial velocity amplitude (top panel) and phase (bottom panel) variations for the lines of \ion{Pr}{ii} (open diamonds), \ion{Pr}{iii} (filled diamonds), \ion{Nd}{ii} (open triangles), \ion{Nd}{iii} (filled triangles), and the H$\alpha$ core (filled circles) as a function of the optical depth
in the atmosphere of the roAp star HD~24712 with the stratified distribution of Pr and Nd.}
\label{puls}
\end{center}
\end{figure}

 The knowledge of the Pr abundance distribution is important for a study of atmospheric pulsations. Fig.\,\ref{puls} shows a distribution of 
the pulsation RV amplitudes and phases of the Pr lines as well as of the lines of \ion{Nd}{ii} and \ion{Nd}{iii}\, and of the H$\alpha$ core  in the atmosphere 
of HD~24712.
The RV data are taken from Ryabchikova et~al. (\cite{Ryabchikova2007a}, Online Table~4).
We use the average optical depths of line formation which are calculated with the element abundance distributions obtained in this study for the Pr and 
by Mashonkina et~al. (\cite{mash_nd_ap}) for the Nd. We follow the formalism suggested by Achmad ~et~al. (\cite{achmad91}) with using  the contribution 
function to the emergent line radiation. The computed values $\log \tau_{5000}^{l}$ are given in Table~\ref{HD24712} (column 12). For comparison, 
we present there (column 13) also the average optical depths of line formation based on 
the contribution function to the emergent total (line + continuum) radiation, $\log \tau_{5000}^{l+c}$. It was emphasized by Achmad ~et~al. that the first 
approach provides more realistic depth of line formation, 
in particular, for weak lines, for which the continuum contribution shifts an average depth formation downward. 
This effect becomes much more important for the case of stratified atmosphere. 
When the contribution function to the total radiation is used, the formation depth of most \ion{Pr}{ii}\ lines is shifted well below the enhanced Pr abundance 
layer. 
 The non-LTE formation depths for the H$\alpha$ core have been computed 
according to Mashonkina et~al. (\cite{MZG07}). 
 It can be seen from Fig.\,\ref{puls}, that with the stratified Pr and Nd abundance distributions found based on non-LTE line formation, we get a consistent picture
typical for running pulsation wave where the amplitudes and phases grow towards upper layers.
A gap of $\sim$0.5~dex at $\log\tau_{5000}\sim-4.5$ may be caused by the uncertainties of the modelling.

\subsection{The uncertainties of Pr stratification analysis}
\label{uncertainty}


In this subsection, we discuss the effects due to the uncertainties of atomic parameters
and due to ignoring magnetic intensification of spectral lines.

{\it Photoionization cross-sections.} The statistical equilibrium of \ion{Pr}{ii/iii}\, in the stratified atmosphere is mainly defined by enhanced photoionization of 
\ion{Pr}{ii}. We perform test calculations using various photoionization cross-sections. When the hydrogenous photoionization cross-sections are increased by a 
factor of 30 the overionization of \ion{Pr}{ii}\, is amplified and $\Delta_{\rm NLTE}$ increases by 0.03~dex to 0.05~dex for different \ion{Pr}{ii}\, lines.
The uncertainty of
photoionization cross-sections is much less important for \ion{Pr}{iii}. 
Reducing the hydrogenous photoionization cross-sections by a factor 
of 100 has much larger effect on both the \ion{Pr}{ii} and \ion{Pr}{iii}\, lines. 
In this case, the layer of enhanced Pr has to be shifted outward by
$\Delta\log\tau_{5000} \simeq$ 0.5 to agree the abundances of Pr from two ionization stages. The obtained Pr abundance distribution is shown by dashed curve in 
the top panel of Fig.\,\ref{bf_layer}. We emphasize that even with the lowest photoionization
cross-sections
the ionization equilibrium \ion{Pr}{ii}/\ion{Pr}{iii}\, deviates significantly
from the TE one and non-LTE removes approximately 1.5~dex of the difference
between the Pr abundances derived from the \ion{Pr}{ii} and \ion{Pr}{iii}\,
lines under the LTE assumption. 
As was discussed in our earlier paper (Mashonkina et~al. \cite{mash_nd_ap}),
the adopted hydrogenic approximation gives, probably, the low limit for the
photoionization cross-sections for the levels in the rare-earth elements. So, we under- rather than overestimate non-LTE effects for \ion{Pr}{ii/iii}\, in our
calculations.

{\it Collision rates.} In SE computations, we assume an effective collision strength to be equal $\Upsilon$ = 1 for every forbidden transition. This may be incorrect. 
The $R-$matrix method calculations for electron impact excitation in \ion{Ca}{ii} (Mel\'endez et~al. \cite{ca2_bautista}) show that 
$\Upsilon$ depends on the transition energy separation. For example, at electron temperatures we are concerned with, $\Upsilon >$ 30 for the 
forbidden transitions with $\Delta E_{ij} <$ 0.1~eV, while $\Upsilon \le$ 0.1 if $\Delta E_{ij} >$ 10~eV. For the fine-structure transitions 
in \ion{Fe}{ii}, the $R-$matrix method predictions of Ramsbottom et~al. (\cite{fe2_coll}) give $\Upsilon \ge$ 5. We perform non-LTE calculations for 
\ion{Pr}{ii/iii}\, assuming that for the forbidden transitions, $\Upsilon$ depends on the transition energy separation as follows: $\log\Upsilon$ = 1.5 for $\Delta E_{ij} <$ 0.1~eV, $\log\Upsilon = -0.38 \Delta E_{ij} + 1.54$ for $\Delta E_{ij} =$ 0.1~eV -- 4~eV, $\log\Upsilon = -0.17 \Delta E_{ij} + 0.67$ for $\Delta E_{ij} =$ 4~eV -- 10~eV, and $\log\Upsilon$ = --1 for $\Delta E_{ij} \ge$ 10~eV. These approximations are based on an extensive set of the data of Mel\'endez et~al. (\cite{ca2_bautista}) for the forbidden transitions in \ion{Ca}{ii}.
With stronger collisional coupling the low-excitation levels to the \ion{Pr}{ii}\, ground state the departures from LTE are weakened for the 
\ion{Pr}{ii}\, transitions arising from the levels with \Eexc\, $<$ 0.8~eV, and they are strengthened for the remaining transitions in 
\ion{Pr}{ii}\, and for all transitions in \ion{Pr}{iii}. The Pr abundance distribution obtained from these test calculations is shown by dotted curve in the top panel of Fig.\,\ref{bf_layer}. We find that no revision is required outside $\log\tau_{5000}$ = --6.5 and inside $\log\tau_{5000}$ = --4 compared to the distribution found in Sect.\,\ref{pr_hd24712}, while the praseodymium abundance is decreased by up to 0.2~dex in between.  

{\it Magnetic field.} The magnetic field effects are taken into account in this study approximately by introducing a 
pseudo-microturbulence. First, this works only for strong spectral lines. Second, magnetic field and microturbulence affect the saturated 
line profile in different ways. Magnetic field splits the line into the Zeeman components making the line less deep 
(magnetic desaturation) and broader, while microturbulence does not
affect the line depth, but increases the half-width of the line profile. As was shown in Sect.~\ref{star_uniform}, taking into account Zeeman splitting 
produces small changes, by 0.16~dex, on average, in the derived Pr abundances compared to those determined without magnetic field. Such an effect cannot 
influence our conclusion on stratified distribution of Pr in the atmosphere of HD\,24712. Zeeman splitting may also result in different line formation depths 
compared to that for turbulently broadened lines and, therefore, in a different location of the Pr enriched layer.  We estimate this effect using the strongest 
\ion{Pr}{iii} $\lambda$5300\AA\ line. 
Zeeman splitting of this line is represented by the three components.
From comparison of the computed  
single line with \Vmic\ = 1.05~\kms\ with the triplet with zero microturbulence we deduce that the magnetic desaturation may lead to only a 0.2~dex 
decrease of the Pr abundance in the layer and in the shift of Pr abundance profile downward, by 0.35~dex in $\log\tau_{5000}$ scale.

 Thus, the uncertainty of the used atomic data on photoionization cross-sections and electron impact excitation cross-sections and ignoring  
the magnetic field in non-LTE analysis cannot change our conclusion as regards an existence of inhomogeneous distribution of Pr in the atmosphere of HD~24712.

\section{Conclusions}\label{end}

In this study, we first presented a comprehensive model atom for \ion{Pr}{ii/iii}\, based on the measured and the calculated energy levels. Non-LTE line formation 
for an extended list of \ion{Pr}{ii} and \ion{Pr}{iii} lines was considered for the Sun and for the
temperatures characteristic of A type stars, $\Teff$ = 7250~K - 9500~ K. At $\Teff \le$ 8000~K, \ion{Pr}{ii}\, represents the majority of the element and the 
departures from LTE for the \ion{Pr}{ii} lines are small. They are mainly caused by a deviation of the line source function from the Planck function and may 
result in the non-LTE abundance corrections of different sign for various lines. Overionization of \ion{Pr}{ii} at $\Teff \ge$ 8500~K leads to depleted total 
absorption in the \ion{Pr}{ii} lines and positive $\Delta_{\rm NLTE}$ that grows with temperature. Non-LTE leads to strengthening the \ion{Pr}{iii} lines and 
negative abundance corrections over the whole range of stellar parameters. As \ion{Pr}{iii} becomes the majority species, $\Delta_{\rm NLTE}$ of the \ion{Pr}{iii} 
lines decreases in absolute value. 

Using the modern laboratory data for oscillator strengths of the \ion{Pr}{ii} lines, improved partition function of \ion{Pr}{ii}\, based on the laboratory and 
the calculated energy levels, we determine the non-LTE abundance of praseodymium in the solar atmosphere, 
$\log\rm {(Pr/H)}_\odot=-11.15\pm0.08$ and $\log\rm {(Pr/H)}_\odot=-11.10\pm0.08$, for the {\sc ATLAS9} and the Holweger \& M\"uller (\cite{HM}) models, respectively. 
The Pr abundance based on the theoretical model is 0.05~dex to 0.10~dex larger compared to the  meteoritic value recommended by Grevesse et~al. (\cite{met96}), Palme \& Jones (\cite{Palme2005}),  Lodders (\cite{Lod}), or Asplund et~al. (\cite{met05}), but the deviation is still within the error bars. 

For the roAp star HD\,24712, the element abundances from two ionization stages, \ion{Pr}{ii} and \ion{Pr}{iii}, reveal a discrepancy of two orders of magnitude if the atmosphere is assumed to be chemically homogeneous. Neither non-LTE, nor magnetic effects can be responsible for such abundance anomaly.  

 Introducing the layer with a strongly enhanced Pr abundance in the outer atmosphere provides a natural possibility to describe the lines of both ionization stages, \ion{Pr}{ii} and 
 \ion{Pr}{iii}, for the single element distribution. We find that the required Pr overabundance in the layer is [Pr/H] $\ge$ 4 at $\log \tau_{5000} < -4$. 
The obtained praseodymium and neodymium stratifications provide a possibility to explain the distributions of the pulsational characteristics 
(radial velocity amplitudes and phases) over the significant part of the atmosphere of HD~24712, where the lines of these elements as well as the core of the H$\alpha$ line are formed.

Similar abundance profiles found empirically for the Nd (Mashonkina et~al. \cite{mash_nd_ap}) and the Pr in the atmosphere of HD\,24712 point to some common physical mechanisms producing accumulation of these two elements in the uppermost atmospheric layers. The main mechanism can be radiatively driven diffusion. 
The theoretical diffusion calculations of LeBlanc \& Monin (\cite{leblanc}) for Cr and Fe and of Alecian \& Stift (\cite{alecian2007}) for Mg, Si, Ca, Ti, Fe predict the existence of the abundance jumps up to several orders of magnitude in stable stellar atmospheres with effective temperatures characteristic of Ap stars. Similar abundance profiles are derived empirically for these elements in the atmospheres of roAp stars (Ryabchikova \cite{RYABCHIK2008}). Opposite to REEs, Mg to Fe are accumulated in the lower atmosphere.
Due to difficulties with the atomic data
no diffusion calculations have been performed for the REEs to check the validity of the empirical distributions, but this situation is improving and 
we hope to see first REE diffusion calculations in the future. 

We are planning to extend non-LTE line formation study to other REE which reveal a discrepancy between the element abundances derived from the lines of 
two ionization stages in the roAp stars. As was shown for Nd by Mashonkina et~al. (\cite{mash_nd_ap}) and for Pr in the present work, the departures from 
LTE for the lines of the first and the second ions are of the opposite sign, and they are large if the element is concentrated in the uppermost atmospheric 
layers where collisions are inefficient to establish thermodynamic equilibrium. In such a case, a stratification analysis for REE has to be performed based on non-LTE line formation. 


\begin{acknowledgements}
The authors are grateful to J.-F. Wyart for help in calculations of \ion{Pr}{ii}\, and \ion{Pr}{iii} spectra.
This research was supported by the Russian Foundation for Basic
Research with grant 08-02-00469-a, by the Presidium RAS Programme ``Origin and evolution
of stars and galaxies'', and by the Leading Scientific School grant 4224.2008.2. TR also acknowledges a partial support from the Austrian Science Fund (FWF-P17580N2).
\end{acknowledgements}

\Online



%

\begin{thebibliography}{99}
\bibitem[1991]{achmad91}
Achmad, I., de Jager, C., \& Nieuwenhuijzen, H. 1991, A\&A, 250, 445
\bibitem[2007]{alecian2007}
Alecian, G. \& Stift, M. J. 2007, A\&A, 475, 659
\bibitem[1998]{allard98}
Allard, N. F., Drira, I., Gerbaldi, M., et al. 1998, A\&A, 335, 1124
\bibitem[1989]{AG}
Anders, E., Grevesse, N. 1989, Geoch. \& Cosmochim Acta 53, 197
\bibitem[1995]{omara_sp}
Anstee, S.D. \& O'Mara, B.J. 1995, MNRAS, 276, 859
\bibitem[2005]{met05} 
Asplund, M., Grevesse, N., \& Sauval, A.J. 2005, ASP Conf. Ser., 336, 25
\bibitem[1991]{BKK}
Bard, A., Kock, A., \& Kock, M. 1991, \aap, 248, 315
\bibitem[1979]{B79}
Bi\'emont, E., Grevesse, N., \& Hauge, \O. 1979, Solar Phys., 61, 17
\bibitem[2001]{B2001}
Bi\'emont, E., Garnir, H.P., Palmeri, P., et~al. 2001, Phys.Rev., A64, 2503
\bibitem[2003]{B2003}
Bi\'emont, E., Lefebvre, P.H., Quinet, P., et~al. 2003, European
Phys.J., D27, 33
\bibitem[1971]{Brewer}
Brewer, L. 1971, JOSA, 61, 1666
\bibitem[1985]{detail}
Butler K., \& Giddings J. 1985, Newsletter on the analysis of astronomical spectra No. 9, University of London
\bibitem[1996]{cgm}
Canuto, V. M., Goldman, I., \& Mazzitelli, I. 1996, ApJ, 473, 550
\bibitem[2001]{cast_kur2001}
Castelli, F. \& Kurucz, R. 2001, A\&A, 372, 260
\bibitem[1981]{cowan}
Cowan, R.D. 1981, The Theory of Atomic Structure and Spectra, Univ.of
California Press. Berkeley. California. USA
\bibitem[1998]{cowley98}
Cowley, C.R. \& Bord, D.J. 1998, in The scientific impact of the Goddard High Resolution Spectrograph. 
ASP Conf.Ser. 143, 346
\bibitem[2000]{cowley2000}
Cowley, C.R., Ryabchikova, T., Kupka, F., et~al. 2000, MNRAS, 317, 299
\bibitem[1961]{Draw}
Drawin, H.-W. 1961, Z.Physik 164, 513
\bibitem[1997]{Fuhr1}
Fuhrmann, K., Pfeiffer, M., Frank, C., et al. 1997, A\&A, 323, 909
\bibitem[2001]{furman}
Furman, B., Stefanska, D., Stachowska, E., et~al. 2001, European Phys.J. D17, 275
\bibitem[2000]{gelbman}
Gelbmann, M., Ryabchikova, T.A., Weiss, W.W., et~al. 2000, A\&A, 356, 200
\bibitem[1989a]{ginibre1}
Ginibre, A. 1989a, Physica Scripta, 39, 694
\bibitem[1989b]{ginibre2}
Ginibre, A. 1989b, Physica Scripta, 39, 710
\bibitem[1990]{Ginibre1990}
 Ginibre, A. 1990, Atomic Data and Nuclear Data Tables, 44, 1
\bibitem[1996]{met96} 
Grevesse, N., Noels, A., \& Sauval, A.J. 1996, ASP Conf.Ser., 99, 117
\bibitem[2004]{grupp}
Grupp, F. 2004, A\&A, 420, 289
\bibitem[2002]{sun_cgm}
Heiter, U., Kupka, F., van 't Veer-Menneret, C., et~al. 2002, A\&A, 392, 619
\bibitem[1974]{HM}
Holweger, H., \& M\"uller, E.A. 1974, Solar Phys., 39, 19
\bibitem[2001]{ILW}
Ivarsson, S., Litz\'en, U., \& Wahlgren, G. 2001, Phys. Scr., 64, 455
\bibitem[2003]{ILW2003}
Ivarsson, S., Wahlgren, G., \& Ludwig, H.-G. 2003, BAAS, 35, 1421
\bibitem[2003]{Kato03}
Kato, K. 2003. PASJ, 55, 1133
\bibitem[2003]{K03}
Kochukhov, O. 2003, A\&A, 404, 669
\bibitem[2007]{synth3}
Kochukhov, O. 2007, in \textit{Physics of Magnetic Stars}, eds. I.I. Romanyuk and D. O. Kudryavtsev, Nizhnij Arkhyz, p. 109
\bibitem[2005]{Ketal05}
Kochukhov, O., Khan, S., \& Shulyak, D. 2005, A\&A, 433, 671
\bibitem[1999]{vald} Kupka, F., Piskunov, N., Ryabchikova, T.A., et~al. 1999, A\&AS 138, 119
\bibitem[1994a]{Kur94} Kurucz, R. L. 1994a, Opacities for Stellar Atmospheres. CD-ROM No. 2-8. Cambridge, Mass
\bibitem[1994b]{Kur94a} Kurucz, R. L. 1994b, SYNTHE Spectrum Synthesis Programs and Line Data. CD-ROM No. 18. Cambridge, Mass
\bibitem[1995]{Kur94b} Kurucz, R. L. \&  Bell, B. 1995, Atomic Line Data. Kurucz CD-ROM No. 23. Cambridge, Mass
\bibitem[1984]{NSO}
Kurucz, R. L., Furenlid, I., Brault, J., \& Testerman, L. 1984, NSO Atlas No. 1: Solar Flux Atlas from 296 to 1300 nm, Sunspot, NSO
\bibitem[1976]{LW}
Lage, C.S. \& Whaling, W. 1976, JQSRT, 16, 537
\bibitem[2004]{leblanc}
LeBlanc, F. \& Monin, D. 2004, The A-Star Puzzle,IAUS 224, eds. J. Zverko, W.W. 
   Weiss, J. \,\v{Z}i\v{z}\v{n}ovsk\'{y} \& S.J. Adelman, 193 
\bibitem[2007]{LCH}
Li, R., Chatelain, R., Holt, R.A., et~al. 2007, Phys. Scr., 76, 577
\bibitem[2003]{Lod}
Lodders, K. 2003, \apj, 591, 1220
\bibitem[1978]{NIST}
Martin, W.C., Zalubas, R., \& Hagan, L. 1978, Atomic energy levels - The Rare Earth Elements. NSRDS-NBS 60, U.S. Gov. Print. Off., Washington. 1978.
\bibitem[2008]{ml_eas}
Mashonkina, L. 2008, in Non-LTE line formation for trace elements in stellar atmospheres, 
Eds. R. Monier, B. Smalley, Ph. Stee, and G. Wahlgren, EAS Publ. Ser (in press)
\bibitem[2005]{mash_nd_ap}
Mashonkina, L., Ryabchikova, T.A., \& Ryabtsev, A.N. 2005. A\&A, 441, 309
\bibitem[2008]{MZG07}
Mashonkina, L., Zhao, G., Gehren, T. et~al. 2008, A\&A, 478, 529
\bibitem[1976]{MCS}
Meggers, W.F, Corliss, C.H., \& Scribner, B.F. 1975, NBS Monograph 145, U.S. Gov. Print. Off.,  Washington, D.C.
\bibitem[2007]{ca2_bautista}
Mel\'endez, M., Bautista, M.A., \& Badnell, N.R. 2007, A\&A, 469, 1203
\bibitem[2005]{Palme2005}
 Palme, H., \& Jones, A. 2005, in Meteorites, Comets and Planets: Treatise on Geochemistry, vol. 1. Ed. A. M. Davis, Elsevier Publ., Amsterdam, The Netherlands, 41
\bibitem[2000]{Palmeri}
Palmeri, P., Quinet, P., Fremat, Y., et~al. 2000, ApJS, 129, 367
\bibitem[2007]{fe2_coll}
Ramsbottom, C. A., Hudson, C. E., Norrington, P. H., \& Scott, M. P. 2007, 
A\&A, 475, 765
\bibitem[1991]{Reetz}
Reetz, J. K. 1991, Diploma Thesis, Universit\"at M\"unchen
\bibitem[2008]{RYABCHIK2008}
Ryabchikova, T. 2008, Contr. Astron. Obs. Skalnat\'e Pleso, 38, 257
\bibitem[2008]{Ca-IS}
Ryabchikova, T.A., Kochukhov, O.,  \& Bagnulo, S. 2008, A\&A, 480, 811
\bibitem[1997]{RYABCHIK97}
 Ryabchikova, T.A., Landstreet J.D., Gelbmann M.J., et~al. 1997, A\&A, 327, 1137 
\bibitem[2004]{Ryabchikova2004}
Ryabchikova, T., Nesvacil, N., Weiss, W.W., et~al. 2004, A\&A, 423, 705
\bibitem[2002]{ryabchik02}
Ryabchikova, T., Piskunov, N., Kochukhov, O., et~al. 2002, A\&A, 384, 545
\bibitem[2006]{RRKB06}
Ryabchikova, T., Ryabtsev, A., Kochukhov, O., \& Bagnulo, S. 2006, \aap, 456, 329
\bibitem[2007a]{RYABCHIK07b}
Ryabchikova, T., Sachkov, M., Kochukhov, O., \& Lyashko, D. 2007a, \aap, 473, 907
\bibitem[2007b]{Ryabchikova2007a}
Ryabchikova, T., Sachkov, M., Weiss, W.W., et~al. 2007b, A\&A, 462, 1103
\bibitem[2000]{RSHWH00}
Ryabchikova, T.A., Savanov, I.S., Hatzes, A.P., et~al. 2000, A\&A, 357, 981
\bibitem[2001]{ryab2001}
Ryabchikova, T.A., Savanov, I.S., Malanushenko, V.P., \&
Kudryavtsev, D.O. 2001, Astron. Rep., 45, 382
\bibitem[1991]{rh91} Rybicki, G.B., \& Hummer, D.G. 1991, A\&A, 245, 171
\bibitem[1992]{rh92} Rybicki, G.B., \& Hummer, D.G. 1992, A\&A, 262, 209
\bibitem[2002]{Scholl}
Scholl T., Holt R., Masterman D., et~al. 2002, Can.J.Phys., 80, 713
\bibitem[1966]{ca4226}
Smith, W.W. \& Gallagher, A. 1966, Phys. Rev., 145, 26
\bibitem[1984]{hyd}
Steenbock, W. \& Holweger, H. 1984, A\&A, 130, 319
\bibitem[1994]{Takeda1994}
Takeda, Y. 1994, PASJ, 46, 53
\bibitem[1962]{Reg} van Regemorter, H. 1962, ApJ, 136, 906
\bibitem[2000]{WLN}
Wickliffe, M. E., Lawler, J. E., \& Nave, G. 2000, JQSRT, 66, 363
\end{thebibliography}
\end{document}